\documentclass[prc,twocolumn,nofootinbib]{revtex4}

\usepackage{graphicx}% Include figure files
\usepackage{amssymb,amsmath,amstext,amsthm,amsfonts}
\usepackage[bookmarks,dvips,pdfhighlight=/O,pdfstartview=FitH]{hyperref}
\usepackage{slashed}

\newcommand{\GeV}{\; \mathrm{GeV}}

\newcommand{\cm}{\; \mathrm{cm}}

\newcommand{\ppr}{p'}

\newcommand{\di}{\displaystyle}
\newcommand{\slashp}{\slashed{p}}
\newcommand{\slashq}{\slashed{q}}
\newcommand{\slashppr}{\slashed{p}'}

\newcommand{\fff}{\mathcal{F}}
\newcommand{\ffc}{\mathcal{C}}

\newcommand{\ePunkt}{\, .}

\begin{document}

\title{One pion production in neutrino reactions: including nonresonant background}
\author{O. Lalakulich
\email{Olga.Lalakulich@theo.physik.uni-giessen.de}}
\author{T. Leitner}
\author{O. Buss}
\author{U. Mosel}
\affiliation{Institut f\"ur Theoretische Physik, Universit\"at Giessen, Germany}

\begin{abstract}
We investigate neutrino induced one pion production on nucleons. The elementary neutrino--nucleon cross section
is  calculated as the sum of the leading Delta pole diagram and several background
diagrams obtained within  the nonlinear sigma model.
This approach does not introduce any new adjustable parameters, which allows unambiguous predictions for the observables.
Considering electroproduction experiments as benchmark, the model is shown to be applicable up to pion-nucleon
invariant mass $W<1.4\GeV$ and provides a good accuracy. With respect to the total one pion cross section,
the model predicts the background at the level of $10\%$ for the $p\pi^+$, $30\%$ for $p\pi^0$,
and $50\%$ for $n\pi^+$ final states. The results are compared with experimental data for various differential cross sections.
Distributions with respect to muon-nucleon and muon-pion invariant masses are presented for the first time.
The model describes the data quite well, with the discrepancies being of the same order as those between different data sets.

\end{abstract}

\maketitle

\section{Introduction}

The interest in one pion production in neutrino--nucleus reactions has recently been revived
in view of the current experimental search for neutrino oscillations. The neutrino energy spectra
for the ongoing and coming long baseline neutrino experiments are typically peaked in the GeV region, the
region where one pion production along with the quasielastic scattering gives a major contribution. Besides
being interesting as a separate channel, pion production constitutes a noticeable background for
various processes: the pion can be absorbed in the nucleus and thus mimic a quasielastic event,
in Cherenkov detectors $\pi^0$ can mimic the outgoing electron. Thus, a precise knowledge of
the corresponding cross sections is a prerequisite for the proper interpretation of the experimental data.

Understanding of one pion production includes two aspects: a proper description of the
elementary process on nucleon and a proper treatment of the nuclear correction. Here we will
concentrate on the elementary process.

In electromagnetic processes, the one pion production data, being plotted versus the invariant
mass of the outgoing pion and nucleon, is seen as a series of peaks. This picture was a basis for the so-called
isobar models, in which the intermediate state of the reaction was treated as a baryon resonance.
The first prominent peak was shown to originate  mainly from the Delta [$P_{33}(1232)$] excitation. The second broader
peak receives contributions from the
so-called second  resonance region, which includes $P_{11}(1440)$, $D_{13}(1520)$ and $S_{11}(1535)$ resonances.
In electroproduction the resonance excitations are known to be accompanied by the so-called nonresonant
background, which can also interfere with the resonance contribution. Because the theoretical structure of the 
resonance contributions is known the modern precise experiments on meson
electroproduction allow the separation of these
contributions and the extraction of the information related to the resonances only; see, for example,
\cite{Burkert:2004sk} for a review. This information can be expressed in the form of the quasiexperimental
``data points'' for the  invariant helicity amplitudes that characterize resonances and exclude background.

In neutrino production, the corresponding approach is, first, complicated by the
fact that the cross section contains in addition to the vector current contribution also an axial one and a vector-axial interference contribution so that more resonance properties have to be determined. Second, any such extraction of such properties 
suffers from the absence of precise, high-statistics data.
Here the data were obtained mainly in the 1980s in bubble chamber experiments.  The most relevant  ones are the 
hydrogen and deuterium data from the Argonne National Laboratory(ANL)
and the Brookhaven (BNL) National Laboratory \cite{Radecky:1981fn,Kitagaki:1986ct} which all suffer from low statistics 
(in comparison to the electroproduction data) so that only integrated and single-differential  cross sections were reported. 
An additional experimental problem in these and all other neutrino experiments is that one cannot fix the neutrino energy, but has to deal with broad band neutrino beams. Thus, we are facing the problem of fixing both more complicated background and resonant parts from a very restricted set of
data. 

Within the phenomenological models, the way out of this situation was to presume that
in the $\nu p$ reaction, i.e.\ in the isospin-3/2 channel, there is no background for the $\Delta^{++}$ production (see, for example,
\cite{Rein:1980wg,AlvarezRuso:1998hi,Paschos:2003qr}). This was motivated by, first, the measured $\pi\,N$ invariant mass distribution in this channel and, second, by the absence of a nucleon Born term in this isospin-3/2 channel.
Within this picture (once the vector form factors
of the $\Delta$ production are considered to be fixed  from electroproduction data), one can fit the Delta axial
form factors and use them further for other channels. Recent progress in this direction was achieved by
refitting the vector form factors from the up-to-date electroproduction data on helicity amplitudes
\cite{Lalakulich:2006sw,Hernandez:2007ej,Leitner:2008ue}
and refitting the axial form factors in the combined analysis of the ANL and BNL experiments
\cite{Graczyk:2007bc,Hernandez:2010bx}.

Even if the axial form factors are fitted to describe the data for the $p \pi^+$ final state,
we have to go beyond the isobar concept and include background contributions, when considering
$p\pi^0$ and $n\pi^+$ final states. The simplest argument comes from the experimental observation that the cross sections
for these two final states are approximately equal, while the $\Delta$ contribution alone
gives $\sigma(p \pi^0)/ \sigma(n \pi^+)=2$. The calculated cross sections are also shown to be lower than the experimental data.
Including higher resonances, in particular the three isospin-1/2 states from the second resonance region, increases the cross sections and
improves the situation somewhat, but does not account for the missing strength.
The additional contributions required can be introduced within the assumption $\sigma(p \pi^0)/\sigma(n \pi^+)=1/2$,
(the so  called ``isospin-1/2'' background) \cite{Lalakulich:2006sw}.
A similar philosophy was applied recently in
\cite{Leitner:2008ue}, where the vector part of the background was extracted from electroproduction,
as it is described by the MAID group \cite{Drechsel:2007if}, and then the magnitude of the
background was fitted to the ANL neutrino data.

The obvious way beyond this simplest picture is to treat the background as a sum of Feynman diagrams with a pion and a nucleon in the final state. Progress in this direction was achieved by Sato, Uno, and Lee \cite{Sato:2003rq}, and recently by Hernandez, Nieves, and Valverde \cite{Hernandez:2007qq} and Barbero, L\'opez Castro, and Mariano \cite{Barbero:2008zza}.

In this article we use the model presented in \cite{Hernandez:2007qq} and apply it to electron and neutrino scattering
on nucleons. We consider various kinematic distributions and analyze their sensitivity to the background contributions.
With the model at hand, we are also in
the position to check the phenomenological treatment, used in the Giessen Boltzmann--Uehling--Uhlenbeck (GiBUU)
transport model  \cite{gibuu}.  This model describes nucleon-, nucleus-,  pion-, and electron- induced collisions from
some hundred MeV up to hundreds of GeV within one unified framework.
Recently, neutrino-induced interactions were
also implemented for the energies up to few GeV with the results presented in
\cite{Leitner:2006ww,Leitner:2006sp,Leitner:2008ue,Leitner:2008wx,Leitner:2009zz}.
The code is written in modular FORTRAN  and is available for download as an open source  \cite{gibuu}.

All current neutrino-nucleon investigations concentrate on reproducing a limited number of distributions,
which include inclusive cross section, $Q^2$, and $W(N\pi)$  distribution
\cite{Rein:1980wg,AlvarezRuso:1998hi,Leitner:2008ue,Sato:2003rq,Paschos:2003qr,Ahmad:2006cy,Hernandez:2007qq,Graczyk:2007bc,Praet:2008yn}.
At the same time, experimentally available data on distributions on the muon-pion and muon-nucleon
invariant masses  \cite{Radecky:1981fn,Kitagaki:1986ct}, which restrict the dynamics of any model even further,  are ignored.
We concentrate on interactions with nucleon targets and aim at reproducing those distributions.
Interactions with nuclei will be discussed in a forthcoming publication and are not considered in this work.

The article is organized as follows. First, we give a short description of the model used. Then we discuss the
electron interactions in the Delta resonance region, producing $p \pi^0$ and $n \pi^+$ final states.
We especially consider the resonance-background interference and possibility to introduce a simplified
description of the background similar to the one discussed earlier \cite{Leitner:2008ue}. Afterward,
results for neutrino interactions are presented and compared with the available experimental data.

\section{Short description of the model  \label{model}}

In this section we give details about the model we use for the  lepton-nucleon interactions.
We are studying the process of one pion production in lepton interactions with nucleons, i.e.
\[
l(k_\mu) N (p_\mu) \to l(k'{}_\mu) N(p'{}_\mu) \pi(p_\mu)
\]
for various isospin final states.

The 5-fold cross section for one pion production is given by
\begin{equation}
\begin{array}{r}  \di
\frac{d\sigma}{dE' d\cos\theta dE_{\pi} d\cos\theta_\pi d\phi_\pi}
= \frac{|M^2|}{4\sqrt{(pk)^2 - m_N^2 m_l^2}} \times
\\[2mm]
\di \times  \frac{1}{(2\pi)^4} \frac{|\vec{k'}| \, |\vec{p_\pi}|}{8 E'_{p}}
\cdot \delta(E+E_p-E'-E_\pi-E'_p) \ .
\label{5-fold}
\end{array}
\end{equation}
The dynamics of the interaction is encoded in the matrix elements for electromagnetic
(EM) and charged current (CC) interactions
\[
|M^2|=C_{EM,CC} L^{\mu\nu} H_{\mu\nu}, \qquad H_{\mu\nu}=j_\mu j^\dagger_\nu \ ,
\]
where
\[
 C_{EM}=\frac{4\pi\alpha_{\scriptstyle QED}}{Q^2} , \quad C_{CC}=\frac{G_F}{2}\cos\theta \ .
\]

The calculation of the leptonic tensor $L^{\mu\nu}$ is straightforward and gives the standard result.
The hadronic tensor $H_{\mu\nu}$ reflects the essence of the process taken into account.
The hadronic current $j_\nu$ varies from model to model and can include various contributions.

\begin{figure}[ht]
\includegraphics[width=\columnwidth]{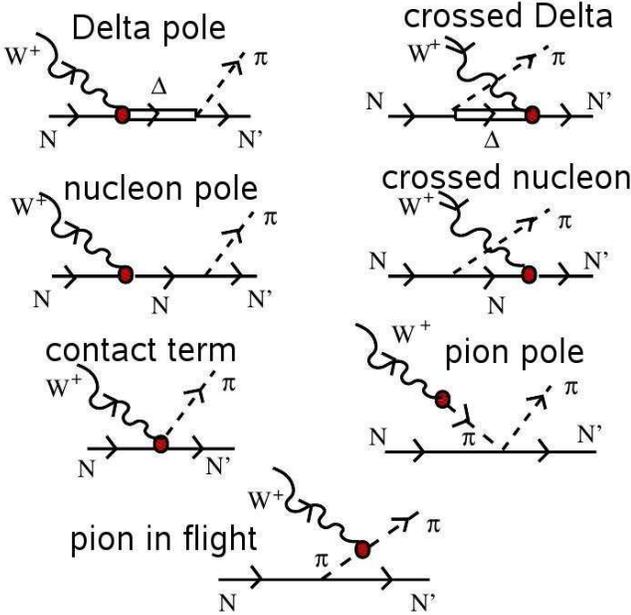}
\caption{Diagrams representing the $\Delta$ pole and background contributions to the one pion production in
weak charged current scattering on the nucleon \cite{Hernandez:2007qq}.}
\label{fig:diagrams}
\end{figure}

Several authors have proposed to describe the current as a coherent sum of several diagrams
\cite{Gil:1997bm,Sato:2003rq,Hernandez:2007qq}:
Delta pole (Dp), crossed Delta pole (cDp), nucleon pole (Np), crossed nucleon pole (cNp), contact term (CT),
pion pole (pp), and pion in flight (pF),
\[
j=j_{Dp}+j_{cDp}+j_{Np}+j_{cNp}+j_{CT}+j_{pp}+j_{pF}
\]
The diagrams considered are shown in Fig.~\ref{fig:diagrams}.

The progress in understanding the background can, however, be only  achieved,
if the new vertices introduced are considered as known and do not include adjustable parameters.

Hernandez et al.\ \cite{Hernandez:2007qq} (from now on called the HNV model) have proposed to use the
vertices predicted by an effective Lagrangian of the $SU(2)$ nonlinear $\sigma$-model.

The details of the model and the amplitudes of the diagrams are given in \cite{Hernandez:2007qq},  and we repeat them here
only for convenience:

%%% Dp
\begin{equation}
\begin{array}{ll}
j_{Dp}^\mu= & \di
  i C^{Dp} \frac{f^*}{m_\pi} \cos\theta_C \frac{p_{\pi}^\alpha}{p_\Delta^2 - M_\Delta^2 + i M_\Delta \Gamma_\Delta} \times
\\[2mm]
  &   \di \times \bar{u}(\vec{p}') S_{\alpha\beta}(p+q) \Gamma_{3/2+}^{\beta\mu}(p,q) u (\vec{p})
\end{array}
\label{Dp}
\end{equation}

%%% cDp
\begin{equation}
\begin{array}{ll}
j_{cDp}^\mu= & \di
  i C^{cDp} \frac{f^*}{m_\pi} \cos\theta_C \frac{p_{\pi}^\beta}{p_\Delta^2 - M_\Delta^2 + i M_\Delta \Gamma_\Delta} \times
\\[2mm]
  &   \di \times \bar{u}(\vec{p}') \gamma^0 [\Gamma_{3/2+}^{\alpha\mu}(p',-q)]^\dagger \gamma^0 S_{\alpha\beta}(p'-q) u(\vec{p})
\end{array}
\label{cDp}
\end{equation}

%%% Np
\begin{equation}
\begin{array}{ll}
j_{Np}^\mu= &  \di
  -i C^{Np} \frac{g_A}{2f_\pi} \cos\theta_C \bar{u}(\vec{p}') \slashed{p}_\pi \gamma_5 \times
\\[2mm]
  &   \di \times \frac{\slashp+\slashq+M_N}{(p+q)^2-M_N^2+i\varepsilon} \left[ V_N^\mu - A_N^\mu \right] u(\vec{p})
\end{array}
\label{Np}
\end{equation}

%%% cNp
\begin{equation}
\begin{array}{ll}   \di
j_{cNp}^\mu= &  \di
  -i C^{cNp} \frac{g_A}{2f_\pi} \cos\theta_C \bar{u}(\vec{p}') \left[ V_N^\mu - A_N^\mu \right] \times
\\[2mm]
  &   \di \times \frac{\slashppr-\slashq+M_N}{(p'-q)^2-M_N^2+i\varepsilon} \slashed{k}_\pi \gamma_5 u(\vec{p})
\end{array}
\label{cNp}
\end{equation}

%%% CT
\begin{equation}
\begin{array}{ll}   \di
j_{CT}^\mu=  & \di
  -i C^{CT} \frac{1}{\sqrt{2} f_\pi} \cos\theta_C \bar{u}(\vec{p}') \gamma^\mu \times
\\[2mm]
  &   \di \times \left( g_A F_{CT}^V( Q^2) \gamma_5 - F_\rho((q-p_\pi)^2) \right) u(\vec{p})
\end{array}
\label{CT}
\end{equation}

%%% pp
\begin{equation}
\begin{array}{ll}   \di
j_{pp}^\mu= &  \di
  -i C^{pp} F_\rho((q-p_\pi)^2) \frac{1}{\sqrt{2} f_\pi} \cos\theta_C \times
\\[2mm]
  &   \di \times \bar{u}(\vec{p}') \slashq u(\vec{p})
\end{array}
\label{pp}
\end{equation}

%%% pF
\begin{equation}
\begin{array}{ll}   \di
j_{pF}^\mu= & \di
  -i C^{pF} F_{pF}(Q^2) \frac{g_A}{\sqrt{2} f_\pi} \cos\theta_C  \times
\\[2mm]
  & \di \times  \frac{(2p_\pi-q)^\mu}{(p_\pi-q)^2-m_\pi^2} 2 M_N  \bar{u}(\vec{p}') \gamma_5 u(\vec{p})
\end{array}
\label{pF}
\end{equation}

Here $g_A=1.26$ is the axial nucleon coupling and $f_\pi=0.093 \GeV$ is the pion weak decay constant,
which enter the Lagrangian of the $\sigma$ model.
The currents defined in (\ref{Dp}) --- (\ref{pF}) can be used for electromagnetic and weak processes, provided that the corresponding
form factors and isospin coefficients $C^{Dp,cDp,Np,cNp,CT,pp,pF}$ are given. In the following  we summarize them.

\subsubsection{$Delta$ resonance}

In the HNV model the vertices of $\Delta$ production and decay (which enter the Delta pole and crossed Delta pole diagrams) are treated
on the same theoretical grounds as earlier in \cite{Paschos:2003qr,Ahmad:2006cy,Praet:2008yn}.

The  $\Delta$ production vertex, $\Gamma^{\beta\mu}$, is described as a vertex for the isospin-3/2 resonance production
\begin{equation}
  \Gamma^{\beta \mu }_{3/2+} = \left[ \mathcal{V}^{\beta \mu }_{3/2} - \mathcal{A}^{\beta \mu }_{3/2}\right] \gamma_{5} \ .
\end{equation}
and used in Eqs.~(\ref{Dp}), (\ref{cDp}).
In terms of the phenomenological form factors, the vector part is given by
  \begin{align}
    \mathcal{V}^{\beta \mu }_{3/2} =&
    \frac{\ffc_3^V}{m_N} (g^{\beta \mu} \slashq - q^{\beta} \gamma^{\mu})+
    \frac{\ffc_4^V}{m_N^2} (g^{\beta \mu} q\cdot \ppr - q^{\beta} {\ppr}^{\mu}) \nonumber \\
 & + \frac{\ffc_5^V}{m_N^2} (g^{\beta \mu} q\cdot p - q^{\beta} p^{\mu})
\label{eq:vectorspinthreehalfcurrent}
\end{align}
  and the axial part by
  \begin{align}
    -\mathcal{A}^{\beta \mu }_{3/2} =& \left[\frac{C_3^A}{m_N} (g^{\beta \mu} \slashq - q^{\beta} \gamma^{\mu})+
      \frac{C_4^A}{m_N^2} (g^{\beta \mu} q\cdot \ppr - q^{\beta} {\ppr}^{\mu})  \right. \nonumber \\ & \left.
 +      {C_5^A} g^{\beta \mu}  + \frac{C_6^A}{m_N^2} q^{\beta} q^{\mu}\right] \gamma_{5} \ePunkt
\label{eq:axialspinthreehalfcurrent}
\end{align}
The calligraphic $\ffc_i^V$ stands either for the electromagnetic transition form factors $C^N_i$ with $N=p,n$
or the CC vector form factors $C^V_i$. For the $\Delta$ resonance, they coincide and
--- in line with \cite{Hernandez:2007qq} --- we use the fit of \cite{Lalakulich:2006sw}:
\[
\begin{array}{l} \di
C_3^V(Q^2)=C_3^{(p,n)}(Q^2)=\frac{2.13}{D_V(Q^2)}\cdot \frac1{1+Q^2/4 M_V^2},
\\[3mm] \di
C_4^V(Q^2)=C_4^{(p,n)}(Q^2)=\frac{-1.51}{D_V(Q^2)}\cdot \frac1{1+Q^2/4 M_V^2},
\label{ff-P1232}
\\[3mm] \di
C_5^V(Q^2)=C_5^{(p,n)}(Q^2)=\frac{0.48}{D_V(Q^2)}\cdot \frac1{1+Q^2/0.776 M_V^2} .
\nonumber
\end{array}
\]
The function $D_V(Q^2)=(1+Q^2/M_V^2)^2$ denotes the dipole function with the vector mass parameter $M_V=0.84\GeV$.
(The axial form factors are relevant only for CC interactions.)
Notice, that in general, the currents for
different isospin channels differ from one another by Clebsch--Gordon coefficients.
In the present work these are included in Eqs.~(\ref{Dp}) --- (\ref{pF}) and defined in Table~\ref{tab:ClebGor}.
Thus, the form factors are the same for different final states.

The axial form factors are taken to be the same as in \cite{Hernandez:2007qq},
where $C_5^A(Q^2)$ was fitted to the ANL cross section:
\[
\begin{array}{l} \di
 C_5^A(Q^2)=\frac{0.867}{D_A(Q^2)}\cdot \frac1{1+Q^2/3 M_A^2} \ ,
\\[3mm] \di
 C_4^A(Q^2) = -\frac14 C_5^A(Q^2), \quad C_3^A(Q^2)=0, \quad C_6^A(Q^2)=0,
\end{array}
\]
with $D_A(Q^2)=(1+Q^2/M_A^2)^2$ and $M_A=0.985 \GeV$.

The value $C_5^A(0)=0.867$ obtained in \cite{Hernandez:2007qq} is in contradiction with the
predictions of the off-diagonal Goldberger--Treiman relation, which expresses $C_5^A(0)$ via
the $\Delta N\pi$ coupling constant $f^*$ and gives the value $1.2$. This relation is based on the
partial conservation of axial current (PCAC) hypothesis, which was tested in  several experiments
and is shown to be satisfied with an accuracy not worse than $10\%$ \cite{Hite:1976ff,Kitazawa:1984qn,Narison:1988xi}.
When considering both ANL and BNL data, the recent fit \cite{Hernandez:2010bx} gives $C_5^A(0)=1$,
which is closer to the PCAC prediction, but still is outside the $10\%$ deviation from it.
In \cite{Leitner:2008ue} the ANL data were described with a fit for $C_5^A$, that fulfills PCAC.

\begin{table}[!hbt]
\caption{Clebsch--Gordon coefficients for various final states in EM and CC interactions}
\[
\begin{array}{ccccccc}
\hline
\multicolumn{7}{c}{\mbox{EM}}
\\
\hline
	   & 	p\pi^+  & p \pi^0    &  p \pi^-    &   n\pi^+    &   n\pi^0   &   n\pi^-
\\
\hline		
\mbox{Dp} &	0	& \sqrt{2/3} & \sqrt{1/3}  & -\sqrt{1/3} & \sqrt{2/3} & 0
\\
\mbox{cDp} &    0	& \sqrt{2/3} & -\sqrt{1/3} & \sqrt{1/3} & \sqrt{2/3}  & 0
\\
\mbox{Np} &	0	& \sqrt{1/2} &  1          & 1		& -\sqrt{1/2} & 0
\\
\mbox{cNp} &    0	& \sqrt{1/2} &  1          & 1		& -\sqrt{1/2} & 0
\\
\mbox{CT, pF} &	0	& 0	     &  1	   & -1         & 0           & 0
\\
\mbox{pp} &	0       & 0	     &  0	   & 0          & 0           & 0
\\
\hline
\multicolumn{7}{c}{\mbox{CC}}
\\
\hline
\mbox{Dp} & \sqrt{3}	& -\sqrt{2/3} & \sqrt{1/3}  & \sqrt{1/3} & \sqrt{2/3} & \sqrt{3}
\\
\mbox{cDp} & \sqrt{1/3}	& \sqrt{2/3}  & \sqrt{3}    & \sqrt{3}   & -\sqrt{2/3} & \sqrt{1/3}
\\
\mbox{Np} &  0          & \sqrt{1/2}  & 1           & 1          & -\sqrt{1/2} & 0
\\
\mbox{cNp} & 1          & -\sqrt{1/2} & 0           & 0          & \sqrt{1/2}  & 1
\\
\mbox{CT, pp, pF} &  1          & -\sqrt{2}   & -1          & -1         & \sqrt{2}    & 1
\\
\hline
\end{array}
\]
\label{tab:ClebGor}
\end{table}

The spin-3/2 projector is taken in conventional Rarita-Schwinger form
\begin{equation}
\begin{array}{l}
  S_{\alpha \beta} (p_\Delta) =  - \left(\slashp^{\Delta} + M_\Delta  \right) \nonumber
  \\[2mm] \di \hspace*{7mm}
  \times   \left(  g_{\alpha \beta} - \frac{2}{3} \frac{p^\Delta_\alpha p^\Delta_\beta}{M_\Delta^2}
 		+ \frac{1}{3}   \frac{p^\Delta_{\alpha } \gamma_{\beta} - p^\Delta_{\beta } \gamma_{\alpha}}{M_\Delta }
 		- \frac{1}{3}   \gamma_{\alpha} \gamma_{\beta} \right)
\end{array}
\end{equation}
and used in Eqs.~(\ref{Dp}), (\ref{cDp}).

The coupling  $f^{*}$  of the $\Delta N \pi$ vertex in $j_{Dp}$ and $j_{cDp}$ currents is determined
from the free decay width of the $\Delta$ resonance, $f^*=1.15$ \cite{Hernandez:2007qq}.

\subsubsection{Nucleon}

The vertices with nucleons and pions are described
within the $SU(2)$ nonlinear $\sigma$-model. Within this model  all the vertices are pointlike, including the
coupling $V N N$ of a nucleon to the vector or axial current.

The authors of \cite{Hernandez:2007qq} choose to introduce nucleon form factors to the $V N N$ vertex in a phenomenological way.
All these form factors are considered to be known and can be taken from one of the conventional  parameterizations.

For the nucleon vertex we adopt the standard hadronic current $\mathcal{V}_N^\mu-\mathcal{A}_N^\mu$
\begin{equation}
  \mathcal{V}^{\mu}_{N}=\fff_1 \gamma^\mu  + i \frac{\fff_2}{2 M_N} \sigma^{\mu\alpha} q_\alpha ,
   \label{Vmu}
\end{equation}
and the axial part
\begin{equation}
  -\mathcal{A}^{\mu}_{N}= F_A \gamma^\mu \gamma_5 + \frac{F_P}{M_N} q^\mu \gamma_5 \ .
 \label{Amu}
\end{equation}
Here,  $\fff_i^V$ ($i=1,2$) stands either for the EM nucleon form factors $F^N_i$ with $N=p,n$
or the CC form factors $F_i^V=F_i^p-F_i^n$. The electromagnetic Dirac and Pauli
form factors $F^N_i$  can be rewritten
in terms of Sachs form factors, for which we take the updated BBBA-2007 parametrization
\cite{Bodek:2007vi}.
The axial form factors are relevant only for CC reactions and for them we assume a standard dipole form
with the axial mass determined in~\cite{Kuzmin:2007kr}:
\begin{equation}
\begin{array}{l} \displaystyle
F_A(Q^2)= g_A \left(1+\frac{Q^2}{M_A^2}\right)^{-2}, \quad M_A=0.999 \GeV
\\[2mm]  \displaystyle
F_P(Q^2)=\frac{2 m_N^2}{Q^2+m_{\pi}^2}F_A(Q^2) \ .
\end{array}
\label{nucleonFF}
\end{equation}

\subsubsection{Other diagrams}

As soon as nucleon form factors are introduced in the model, the conservation of the vector current
\begin{equation}
 q^\mu (j_{Dp} + j_{cDp} + j_{Np} + j_{cNp} + j_{CT} + j_{pp} + j_{pF})_\mu \stackrel{!}{=}0
\label{current-conserv}
\end{equation}
is no longer fulfilled.  The way to compensate in Eq.~(\ref{current-conserv}) the nonvanishing terms stemming 
from $j_{Np}$ and $j_{cNp}$ is to introduce the corresponding form factors to $j_{CT}$ and $j_{pF}$. 
%However, the non-vanishing terms in (\ref{current-conserv}) stemming from $j_{Np}$ and $j_{cNp}$ can be compensated
%by those from  $j_{CT}$ and $j_{pF}$ if the corresponding form factors are introduced
%also in the latter two currents.
As outlined in \cite{Hernandez:2007qq}, for charged current processes, the corresponding form factors are the same as the
weak vector nucleon ones. It can be shown, that for electroproduction processes, they also stay the same, i.e.,
\begin{equation}
\begin{array}{l}
F_{CT}^{em}(Q^2)=F_{CT}^{CC}(Q^2)=F_1^{p}(Q^2)-F_1^{n}(Q^2) \ ,
\\[2mm]
F_{pF}^{em}(Q^2)= F_{pF}^{CC}(Q^2)=F_1^{p}(Q^2)-F_1^{n}(Q^2) \ .
\end{array}
\label{CT-pF-FF}
\end{equation}

Another phenomenological factor, introduced in the HNV model, is $F_\rho$,
which accounts for the $\rho$-meson dominance in the axial parts of the CT
and pp currents.

For CC reactions  $F_\rho$ is given by
\[
F_\rho^{CC}(Q^2)=\frac{1}{1+Q^2/m_\rho^2}, \qquad m_\rho=0.77 \GeV \ ,
\]
and for EM interactions it is zero,
\[
F_\rho^{em}(Q^2)=0 \ .
\]

To summarize, the HNV model \cite{Hernandez:2007qq} phenomenologically extends the ``pure'' nonlinear $SU(2)$ model,
but introduces no adjustable parameters. Besides the nucleon and the pion, the model contains only the Delta
resonance and thus is applicable to the region below and slightly above the
Delta peak.

%%%%%%%%%%%%%%%%%%%%%%%%%%%%%%%%%%%%%%%%%%%%%%%%%%%%%%%%%%%%
%%%  electroproduction
%%%%%%%%%%%%%%%%%%%%%%%%%%%%%%%%%%%%%%%%%%%%%%%%%%%%%%%%%%%%

\section{Electroproduction as benchmark for neutrinoproduction  \label{electron} }

In this section, we present the differential cross section results for electron scattering,
with the purpose to check the accuracy of the model  and the range of its applicability.

We consider electrons of energy $E_e=1.884 \GeV$ scattered on protons over the angle $\theta=47.94^\circ$ ($\cos\theta=0.67$)
and calculate the double differential cross section $d\sigma/d\Omega_e dE_e$.
%As expected, the biggest contribution comes from the Delta pole diagram, the others give smaller, but still
%noticeable contributions to the cross section.
%One of the most important ones is the contact term, which provides a sharp increase of the cross section near the threshold.

\begin{figure}[htb]
  \includegraphics[width=\columnwidth]{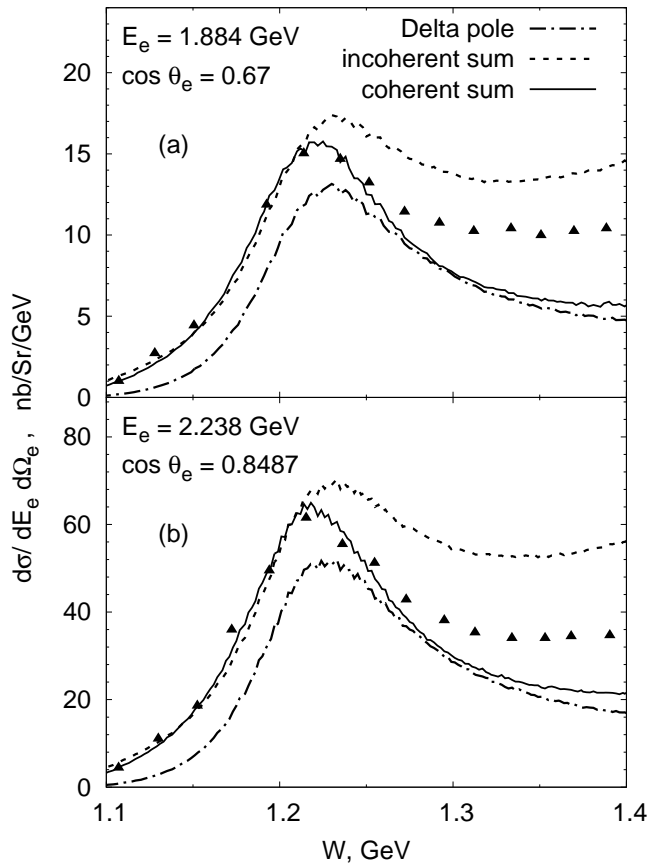}
\caption{Differential cross section for $ep\to e\Delta^+$ at $E_e=1.884\GeV$, $\theta_e=47.94^\circ$ (a)
and at $E_e=2.238\GeV$, $\theta_e=31.93^\circ$ compared to the  inclusive $ep \to e X$ data  \cite{Christy:2007ve}.}
\label{fig:E1884-E2238}
\end{figure}

Fig.~\ref{fig:E1884-E2238}a shows the full model cross section (solid line), obtained as a coherent sum of all the diagrams.
It is compared with JLab electroproduction data~\cite{Christy:2007ve}. Notice, that these data are for the inclusive cross section,
while our curve is for the one pion production, that is for the sum of the $p\pi^0$ and $n \pi^+$ final states, only.
Below and at the $\Delta$ peak, %where other resonances and channels  are known to give negligible contribution,
our calculations agree perfectly with the data. At the same time, the Delta pole diagram alone (dash-dotted line) is noticeably
below the data.
Above the $\Delta$ peak, as expected, the data  lie above our curve, because other resonances
contribute in this region as well as other channels (for example, two pion production and eta production) become kinematically
allowed.
The incoherent sum of all the diagrams (short-dashed line) is also shown in Fig.~\ref{fig:E1884-E2238}. Below
the $\Delta$ peak the interference effects are small, while above the Delta peak the interference is
strong and negative.

The similar picture is also shown in Fig.~\ref{fig:E1884-E2238}b for $E_e=2.238 \GeV$ and $\theta=31.93^\circ$ ($\cos\theta=0.8487$).
The agreement with the data is again very good below and at the $\Delta$ peak.

Conventionally, experimental results for one pion production are shown in the form of the cross
section for virtual photons
\begin{equation}
\frac1{\Gamma_t} \frac{d\sigma}{d \Omega' dE'} = \sigma_T + \varepsilon \sigma_L,
\label{virtual-photon-xsec}
\end{equation}
where $\Gamma_t$  is the flux of the virtual photon field
\[
\Gamma_t =\frac{\alpha_{\scriptstyle QED}}{2\pi^2} \frac{E'}{E_e} \frac{W^2-m_N^2}{2m_N Q^2} \frac{1}{1-\varepsilon},
\]
and $\varepsilon$ is the degree of transverse polarization of the photon
\[
\varepsilon=\left[1+2\left(1+\frac{\nu^2}{Q^2}\right)\tan^2\frac{\theta}{2}\right]^{-1}.
\]
Here $\theta$ is the electron scattering angle, $\nu$ the energy transfer, $Q^2$ the squared momentum transfer,
and $W^2=m_N^2+2m_N\nu-Q^2$ the invariant mass of the final nucleon-pion state.

\begin{figure*}[tbh]
\includegraphics[width=0.7\textwidth]{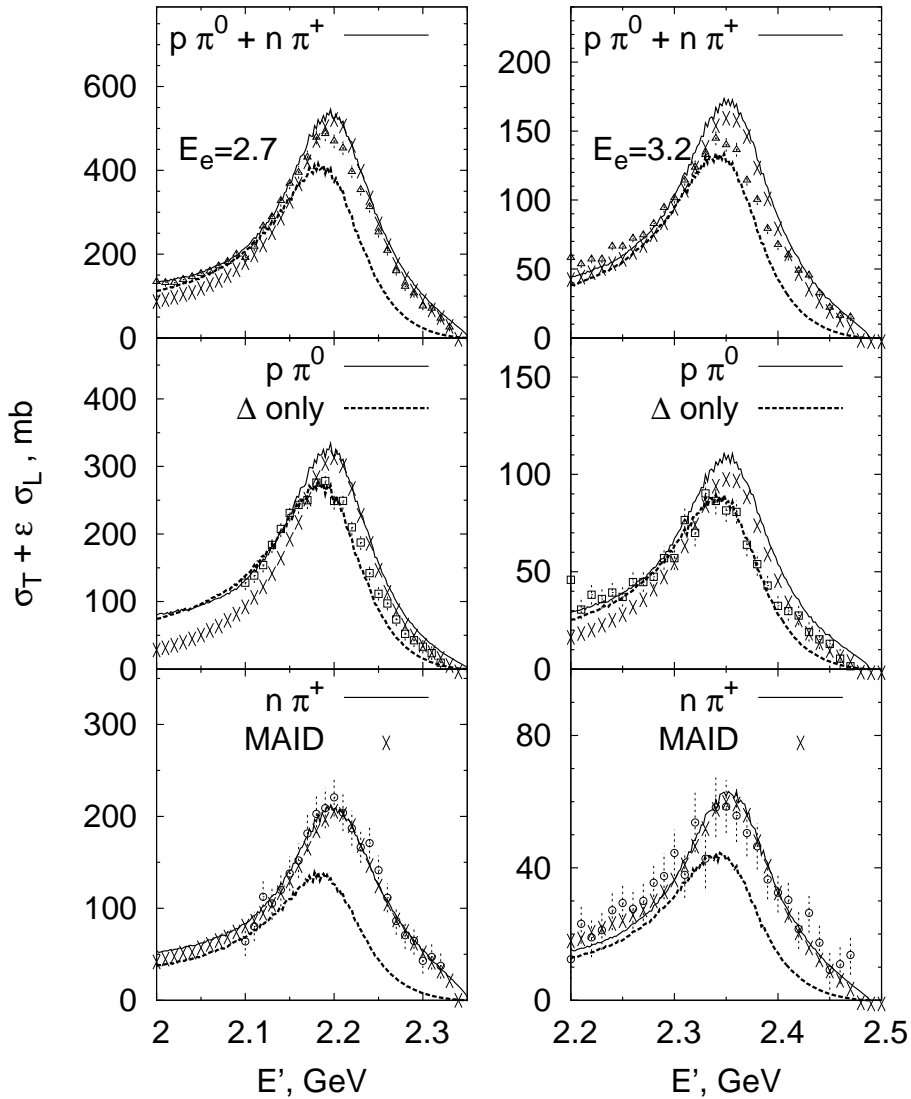}
\caption{Cross section (\ref{virtual-photon-xsec}) for $ep \to e p \pi^0$ (middle panel) and  $ep \to e n \pi^+$ (lower panel)
at  $E_e=2.7\GeV$, $\theta_e=14^\circ$ (left panel)
and $E_e=3.2 \GeV$, $\theta_e=21^\circ$ (right panel) as a function of outgoing electron energy $E'$
compared to the data \cite{Galster:1972rh}.
The predictions of the MAID model are shown as crosses.}
\label{fig:Galster}
\end{figure*}

Our  results are compared to the DESY electron--proton scattering data \cite{Galster:1972rh} in  Fig.~\ref{fig:Galster}
for $p \pi^0$ and $n \pi^+$ final states (middle and lower panels). Two data sets are available:
(1) for electron energy $E_e=2.7 \GeV$ and the scattering  angle of $14^\circ$,  at the $\Delta$ peak position $Q^2=0.35 \GeV^2$;
and  (2) for electron energy $E_e=3.2 \GeV$ and the scattering
angle of $21^\circ$,  at the $\Delta$ peak position $Q^2=1.0 \GeV^2$.
The figure shows the cross section (\ref{virtual-photon-xsec}) versus the outgoing electron energy $E'$ in the $\Delta$ region.
Here the higher $E'$ values corresponds to lower invariant masses $W$.  
We present the full model calculations (solid lines), as well as the contribution of the Delta pole
alone (dashed lines).

The predictions of the MAID model \cite{Drechsel:1998hk,Drechsel:2007if} are also shown as crosses,
which provides us an overall comparison with modern electroproduction data.
The MAID model, developed by Mainz theory group, is a state-of-the-art unitary isobar model
for pion photo- and electroproduction on the nucleon,
which fits more than 70000 data points on 5 and more fold differential cross sections.
For the GiBUU code, it provides the resonance amplitudes for electroproduction. The results of
the MAID model for double differential cross section can be considered as being equivalent
to the data. They can, therefore, serve as a benchmark for our calculation that --- contrary to
MAID --- contains a theoretically well founded description of the background amplitudes.

The full model calculations  show an excellent agreement with the data \cite{Galster:1972rh} for the $n \pi^+$ final state.
At high $E'$, corresponding to the invariant mass region below the Delta resonance, we observe a noticeable increase of the
cross section in comparison with the Delta pole diagram, which significantly improves the agreement with the data.
With decreasing $E'$, the invariant mass $W$ increases, reaching $W=1.29 \GeV$ at the left end of the data points
for $E_e=2.7 \GeV$ and $W=1.35 \GeV$ for $E_e=3.2 \GeV$.
For the $p \pi^0$ final state, our full model, as well as the MAID model, shows a reasonable agreement with the data.

The data are also available for the sum of the final states $p \pi^0 + n \pi^+$
(triangles in the upper panel of Fig.~\ref{fig:Galster}), they agree with our curves up to $W=1.4 \GeV$.
In all cases the full model calculations are very close to the MAID results. Thus, the model
provides the same level of accuracy as the MAID model, which ensures
the applicability of the HNV model to the leptoproduction processes at least up to $W<1.4 \GeV$.

%%%%%%%%%%%%%%%%%%%%%%%%%%%%%%%%%%%%%%%%%%%%%%%%%%%%%%%%%%%%
%%%  neutrino
%%%%%%%%%%%%%%%%%%%%%%%%%%%%%%%%%%%%%%%%%%%%%%%%%%%%%%%%%%%%

\section{Neutrinoproduction \label{neutrino} }

In this section we present our results for neutrinos.

\begin{figure}[hbt]
\includegraphics[width=\columnwidth]{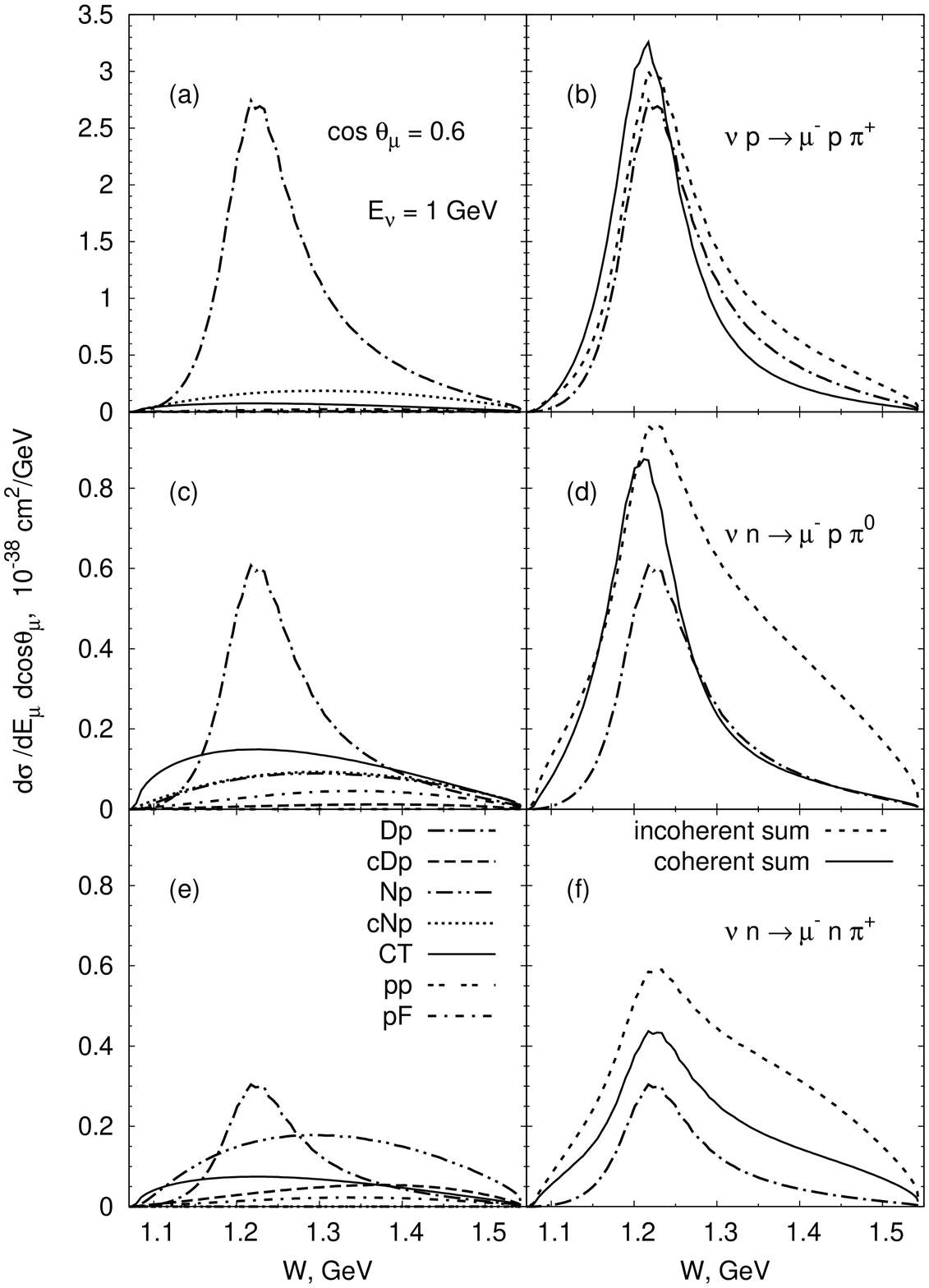}
\caption{Double differential cross section $d\sigma/d E_\mu d \cos\theta_\mu$
for various final states for $E_\nu=1\GeV$, $\cos\theta=0.6$.  Contributions of each diagram (a, c, e),
as well as their coherent and incoherent sums (b, d, f) are shown.
}
\label{fig:pn-E1}
\end{figure}

The double differential cross sections $d\sigma/dE_\mu d\cos\theta_\mu$ for the charged current neutrino reactions
versus the nucleon--pion invariant mass $W$ are presented in Fig.~\ref{fig:pn-E1}  for the incoming
neutrino energy $E_\nu= 1 \; \mathrm{GeV}$ and the muon scattering angle $\cos\theta_\mu=0.6$.
As we already mentioned, the form factors used are taken to be same as in \cite{Hernandez:2007qq}, in particularly we use the same $C_5^A$.
Fig. ~\ref{fig:pn-E1}a shows the contribution of each diagram to the cross section for the $p\pi^+$ final state.
For this channel, the cross sections for the background diagrams are indeed small in comparison with the Delta pole contribution.
Fig. ~\ref{fig:pn-E1}b compares the full model calculation (solid line)
with the Delta pole only (dash-dotted line) and with the incoherent sum  (short dashed line) of all diagrams.
For the kinematics considered, the interference effect is negative above the $\Delta$ peak, and positive below the $\Delta$ peak.

Figs.~\ref{fig:pn-E1}c-f show the same cross section for a neutron target for the
two possible final states, $p \pi^0$ and $n \pi^+$.
As one can see,  the background terms are noticeable in these cases.
The most important contribution is given by the CT diagram, which provides a rather steep rise of the cross
section at low $W$. The Np diagram dominates the background for the $n\pi^+$ channel and, together with the cDp, gives a large
contribution for the $p\pi^0$ one. The way the background appears in reactions with a neutron  target can partly
be  traced to pure isospin relations.  For example, as can easily be deduced from
Table~\ref{tab:ClebGor}, the leading isobar contribution, Dp, for the  $n\pi^+$ channel is 9
times smaller than that for the $p\pi^+$ one.  At the same time, the cDp term, which was very small for the $p\pi^+$,
is 9 times bigger for the $n\pi^+$ and thus becomes noticeable.
>From Figs.~\ref{fig:pn-E1}d,f  one can see, that the interferences are again negative above the $\Delta$ peak,
and small (positive or negative) below the $\Delta$ peak.  The overall increase of the cross section in comparison
with the Delta pole contribution is, as expected, much more significant than for the proton target.

A feature of the HNV model  is that it introduces the background
not only for the $p \pi^0$ and $n \pi^+$ final states, but also for the $p \pi^+$, that is for the isospin-3/2 channel.
We observe, that in this channel the contribution of the background is at the level of  $10\%$,
which agrees  with the result of \cite{Hernandez:2007qq}.
This justifies the neglect of the background in the isospin-3/2 channel as assumed in earlier works
\cite{Lalakulich:2005cs,Ahmad:2006cy,Graczyk:2007bc,Lalakulich:2006sw,Leitner:2008ue}
and explains why they were still successful in describing the data.

%%%%%%%%%%%%%%%%%%%%%%%%%%%%%%%%%%%%%%%%%%%%%%%%%%%%%%%%%%%%
%%%  comparison Leitner09 background
%%%%%%%%%%%%%%%%%%%%%%%%%%%%%%%%%%%%%%%%%%%%%%%%%%%%%%%%%%%%

\section{General features of effective background \label{compareLeitner} }

\begin{figure}[!hbt]
\includegraphics[width=\columnwidth]{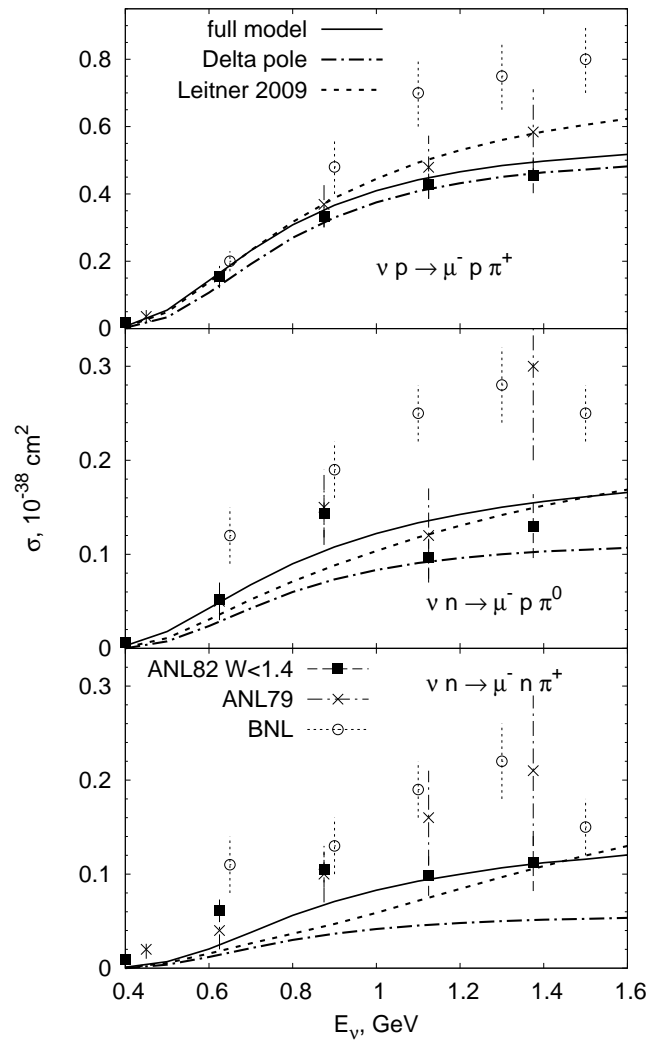}
\caption{The integrated one pion cross section, with kinematical cut $W(N\pi)<1.4 \GeV$ versus neutrino
energy for various final states. The full model calculations (solid line)
are compared with the Delta pole contribution (dash-dotted line) and previous calculations of
Leitner et al~\cite{Leitner:2008ue} (dashed line).}
\label{fig:tot}
\end{figure}

As the next step, in Fig.~\ref{fig:tot} we present our results for the integrated cross section,
with the kinematical cut for the nucleon--pion invariant mass $W(N\pi)<1.4 \GeV$,  versus the neutrino energy.
The calculations are made for various  final states and compared with data and with some previous
theoretical results.

The simplest channel to compare with is the scattering on a proton target, because only one final state, $p \pi^+$, is possible.
The full model calculation (solid line) appears to be slightly above the Delta
pole contribution (dash-dotted line) and coincides with the previous calculation~\cite{Leitner:2008ue}
(dashed line labeled ``Leitner 09'') at small neutrino energies.\footnote{This implementation is available
in the current open-source version of GiBUU \cite{gibuu}.}
With increasing $E_\nu$, the ``Leitner 09'' curve, as expected, increases more steeply than the full model curve because
the calculation~\cite{Leitner:2008ue} was done without any kinematical cut, while our calculation implies $W(N\pi)<1.4\GeV$.
Comparison with Fig.~5 in \cite{Hernandez:2007qq} shows, as expected, that the integrated cross sections is also very close
to the original HNV result.

For the reactions on the neutron two final states, $n \pi^+$ and $p \pi^0$, are possible.
For both of them, the full model cross sections are close to the previous GiBUU results~\cite{Leitner:2008ue},
but have slightly different shapes. Keep in mind, that the $W$ cuts are different.
\footnote{For the $p\pi^0$ channel our result is also in agreement with the original
HNV calculation~\cite{Hernandez:2007qq} (see Fig.~5 there), while for the $n\pi^+$ channel it is noticeably higher
(at $E_\nu=1.6 \GeV$ our result $0.12\cdot 10^{-38} \cm^2$ versus HNV $0.08 \cdot 10^{-38} \cm^2$).
To understand this difference we compared our calculations for each diagram with the corresponding
unpublished results of the HNV group.
%\footnote{kindly provided to us by M. Valverde}.
We found a very good agreement for all diagrams except cDp, which in our calculations
appears to be around 1.7 times smaller. Comparing the Clebsch--Gordon coefficients for various final states
(see Table~\ref{tab:ClebGor}) among themselves, one can notice that the cDp diagram contributes mainly to the $n\pi^+$ channel. Taking into account possible interferences, we attribute the difference
in this channel to this contribution.}

The ANL data can be described quite well for all channels and over the full
energy range. This agreement is trivial for the $p\pi^+$ channel because, as we mentioned before, the axial
form factors were fitted to them. The agreement for the two other channel shows that the model gives a very reasonable
estimate for the integrated background.

\begin{figure}[htb]
\includegraphics[width=\columnwidth]{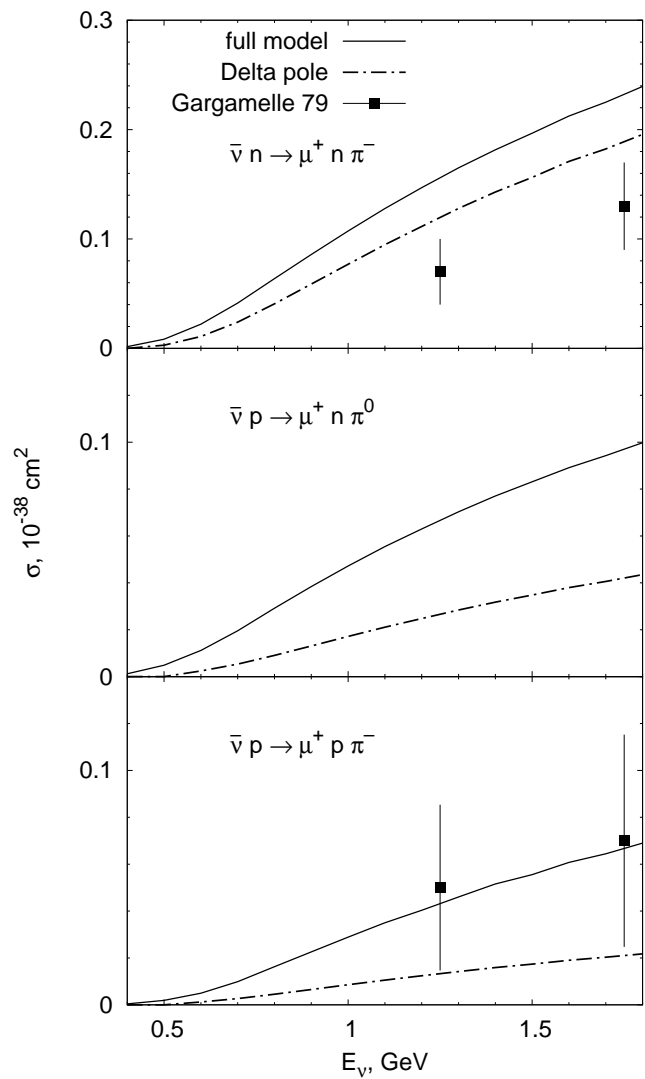}
\caption{The integrated one pion cross section, with kinematical cut $W(N\pi)<1.4 \GeV$ versus the antineutrino
energy for various final states.}
\label{fig:bgr-barnu}
\end{figure}

A few data points from the Gargamelle propane experiment at CERN PS \cite{Bolognese:1979gf} are also available for antineutrino reactions
on neutron $\bar\nu n$ and nucleon $\bar\nu N$, the latter being the sum over proton and neutron targets.
Our full model calculations for the three possible final states, $\bar\nu n \to \mu^+ n \pi^-$, $\bar\nu p \to \mu^+ n \pi^0$,
and $\bar\nu p \to \mu^+ p \pi^-$, are shown in Fig.~\ref{fig:bgr-barnu} and compared to the Gargamelle experiment.
The data on the $p\pi^-$ channel are obtained as
\[
 \sigma(\bar\nu p \to \mu^+ p \pi^-) =  \sigma(\bar\nu N \to \mu^+ N \pi^-) - \sigma(\bar\nu n \to \mu^+ n \pi^-) \ .
\]
The results of our calculations are very close to those presented in the HNV paper~\cite{Hernandez:2007qq}.
The cross section for $n \pi^-$ is overestimated, while that for $p\pi^-$ shows a good agreement with the data.

Comparing the curves for the full model and Delta pole contributions, one can define the effective
background by subtraction:
\[
 \sigma^{\mbox{eff-bgr}} = \sigma^{\mbox{full model}}-\sigma^{\mbox{Delta pole}} .
\]
By definition, $\sigma^{\mbox{eff-bgr}}$ includes interference terms and thus can take
positive or negative values.
This effective background
can be compared with the phenomenological background used in \cite{Leitner:2008ue}.

\begin{figure}[!hbt]
\includegraphics[angle=-90,width=\columnwidth]{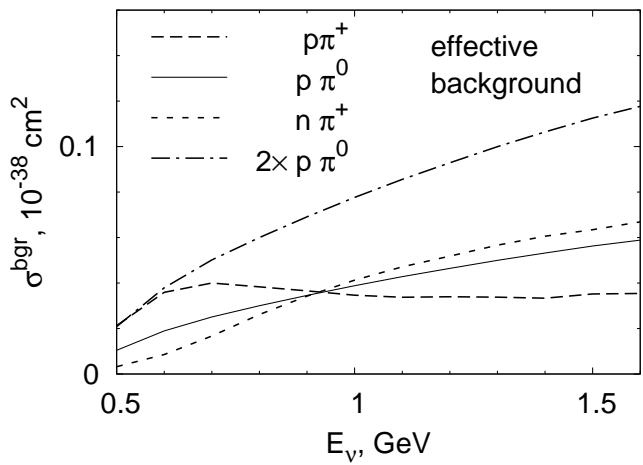}
\caption{The effective background versus the neutrino energy for various final states.}
\label{fig:bgr}
\end{figure}

Fig.~\ref{fig:bgr} shows the effective background for the three final states. For the $p\pi^+$ final state (long--dashed line),
it is at the level of $0.04 \cdot 10^{-38} \cm^2$ for all values of $E_\nu$.
For a neutron target the background is growing with energy taking on about the same values for the $p\pi^0$ (solid line)
and $n\pi^+$ (short--dashed line) final states.

In earlier phenomenological approaches \cite{Lalakulich:2006sw,Leitner:2008ue},
the assumption of the isospin-1/2 background was used,  which presupposes  $ \sigma_{bgr-1/2}^{n \pi^+} = 2 \sigma_{bgr-1/2}^{p \pi^0} $;
this curve  is shown as a dash-dotted line.
Thus, one can conclude, that the effective background does not follow the isospin-1/2 approximation.
Indeed, from the six diagrams (cDp, Np, cNp, CT, pp, pF) directly contributing to the background, only one (Np)
satisfies the isospin-1/2 hypothesis, that is, its Clebsch-Gordon coefficients are related as $C^{Np}_{n\pi^+} = \sqrt{2} C^{Np}_{p\pi^0}$.
For others the corresponding relation is different from $\sqrt{2}$, and interference also plays an important role.

\begin{figure}[!hbt]
\includegraphics[angle=-90,width=\columnwidth]{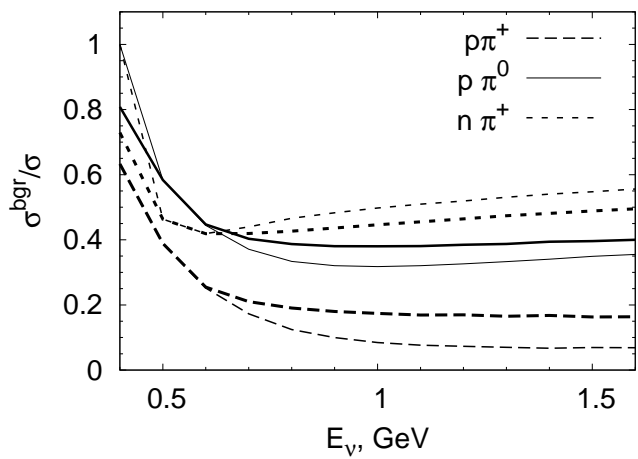}
\caption{The ratios of the effective background to the full model cross section versus the neutrino energy for various
final states for the cuts $W(N\pi)<1.4\GeV$ (thin lines) and $W(N\pi)<1.3\GeV$ (thick lines).}
\label{fig:bgr-percent}
\end{figure}

Fig.~\ref{fig:bgr-percent} shows the ratio of the background to the full model cross section.
To investigate how sensitive this result is to the $W(N\pi)$ cut, we plot two curves for each final state:
with the cut $W(N\pi)<1.4\GeV$ (thin lines) and $W(N\pi)<1.3\GeV$ (thick lines). As one can easily see, the results for
these two cases are very close.
For the $p\pi^+$  channel, the ratio is large for low energy, but steeply falls down and does not exceed $10\%$ for $E_\nu>1\GeV$.
For the neutron target the background is large and is at the level of $35\%$ for the $p\pi^+$ channel and $50\%$ for $n\pi^+$ one.

Recall, that the background seen in neutrino reactions (weak background) include vector, axial and vector-axial-interference parts.
In electron reactions, only the vector part is present (electromagnetic background) and well constrained by data.
In \cite{Leitner:2008ue} it was assumed, that $d\sigma_{bgr}^{A}$ and $d\sigma_{bgr}^{VA}$ have the same functional form as the
vector part:
\begin{equation}
 d\sigma_{bgr}^{V} + d\sigma_{bgr}^{A} + d\sigma_{bgr}^{VA} = (1+ b^{N\pi}) d\sigma_{bgr}^{V} \ .
\label{brg-Tina}
\end{equation}
The vector parts were extracted from data independently for various channels.
The coefficient $b$ was the adjustable parameter fitted to the ANL data under the assumption $b^{n\pi^+}=2b^{p\pi^0}$.

\begin{figure}[!hbt]
\includegraphics[width=\columnwidth]{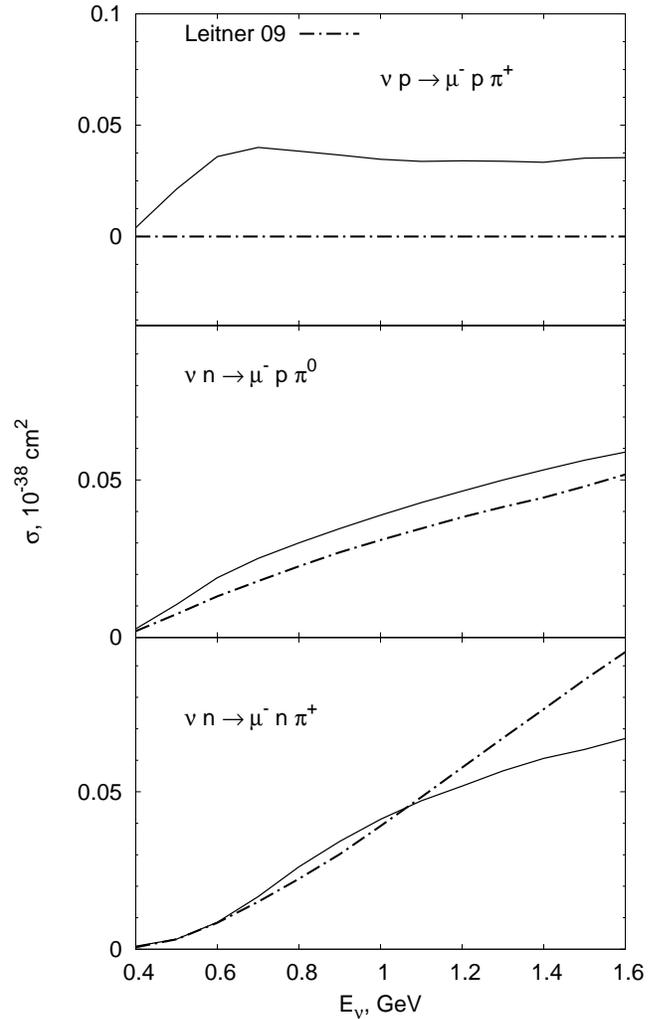}
\caption{The effective background versus the neutrino energy: comparison with the phenomenological
calculations of \cite{Leitner:2008ue}.}
\label{fig:bgr-multi-cutW14}
\end{figure}

Fig.~\ref{fig:bgr-multi-cutW14} compares the effective background of the HNV model (solid lines) with the
phenomenological background (\ref{brg-Tina}) (dash-dotted lines) for the three final states. The agreement between the
curves is  reasonable for all final states up to $E_\nu=1.2 \GeV$, which justifies the model used in \cite{Leitner:2008ue}.

%%%%%%%%%%%%%%%%%%%%%%%%%%%%%%%%%%%%%%%%%%%%%%%%%%%%%%%%%%%%
%%%   Neutrino experiments : ANL
%%%%%%%%%%%%%%%%%%%%%%%%%%%%%%%%%%%%%%%%%%%%%%%%%%%%%%%%%%%%

\section{Comparison with the  ANL  experiment  \label{ANL} }

As we already mentioned, the data on neutrino-nucleon interactions are scarce and come from the late
1970s and early 1980s. In all these experiments wide band neutrino beams were incident on hydrogen and/or deuterium targets.
The most detailed sets of data are provided by the ANL  12-ft  and the BNL 7-ft bubble chambers.
The integrated cross sections from ANL and BNL experiments were already
used in the previous section. However, valuable information comes also from the differential cross sections.

The deuteron effects for relevant neutrino energies were
studied in \cite{AlvarezRuso:1998hi}, where three different wave functions, corresponding to the Hulthen,
Bonn, and Paris NN models, were considered.
It was shown, that these effects depend on the model used and
generally suppress the cross section at $Q^2<0.1\GeV^2$  by no more than $8\%$ for ANL experiment.
At higher $Q^2$ they are practically negligible.
For recent calculations including deuteron effects, see \cite{Graczyk:2009qm,Hernandez:2010bx}.

The GiBUU model implements the Argonne V18 NN potential.
Within this model the deuteron effects are shown \cite{Leitner:2009zz} to introduce only a minor
correction to the ANL $Q^2$ distribution --- the correction is even smaller for BNL because of the higher
neutrino energy.
For invariant mass distributions, which are integrated over $Q^2$, it would be even smaller.
Thus, for the present calculations we neglect the effects of the deuteron structure.

\subsection{Transformation from events to absolute cross section}

Many reaction rates were presented in \cite{Radecky:1981fn} not as absolute cross sections, but as events per
some interval of the measured variable ($Q^2$ or $W$, for example). This is mainly explained by the fact,
that neutrino fluxes are not precisely known, but can only be determined with some accuracy which is hard to estimate.
For the ANL experiment, for example, the flux is calculated from the measured multipion production cross section
on a beryllium target and is given in \cite{Barish:1977qk}.
It is clear, however, that the transformation coefficient $k$ from the number of events per unit energy  to
the absolute cross section is unique for a given experiment and (for a perfect experiment) must be the same
for each reaction channel $ch=(p\pi^+, p\pi^0, n\pi^+)$ for all distributions.

The ANL experiment provides data for the distribution
of observed (also called raw) events $N_{ch}(E_\nu)$ in neutrino energy \cite{Radecky:1981fn}
(see Fig. 7 there),\footnote{Note, that neither ANL nor BNL provide distributions corrected for the experimental
backgrounds, which, ideally, would be the subject of comparison with the theory.}
which is
\begin{equation}
 f_{ch} \cdot  N_{ch}(E_\nu) = \sigma_{tot(ch)}(E_\nu) \cdot flux(E_\nu) \cdot k \  .
\label{k-fit}
\end{equation}
The rate correction coefficients $f_{ch}$ account for experimental backgrounds and losses.
They can be extracted from the summary of rate corrections \cite{Radecky:1981fn} (see Table I there)
or calculated as ratios of corrected to raw events \cite{Radecky:1981fn}:
\begin{equation}
\begin{array}{c}
 f^{ANL}_{p\pi^+}=\frac{1115}{871}=1.280,
\\
 f^{ANL}_{p\pi^0}=\frac{272.8}{202.2}=1.349, \qquad  f_{n\pi^+}^{ANL}=\frac{255.8}{206.2}=1.241 \ .
\end{array}
\label{f_ch}
\end{equation}
With this data in hand and considering the flux  and cross section as experimentally determined,
we are able to calculate the coefficient $k$ for different final states, compare the results
and thus estimate the accuracy intrinsic to the experiment and its consistency.

\begin{figure}[!hbt]
\includegraphics[width=\columnwidth]{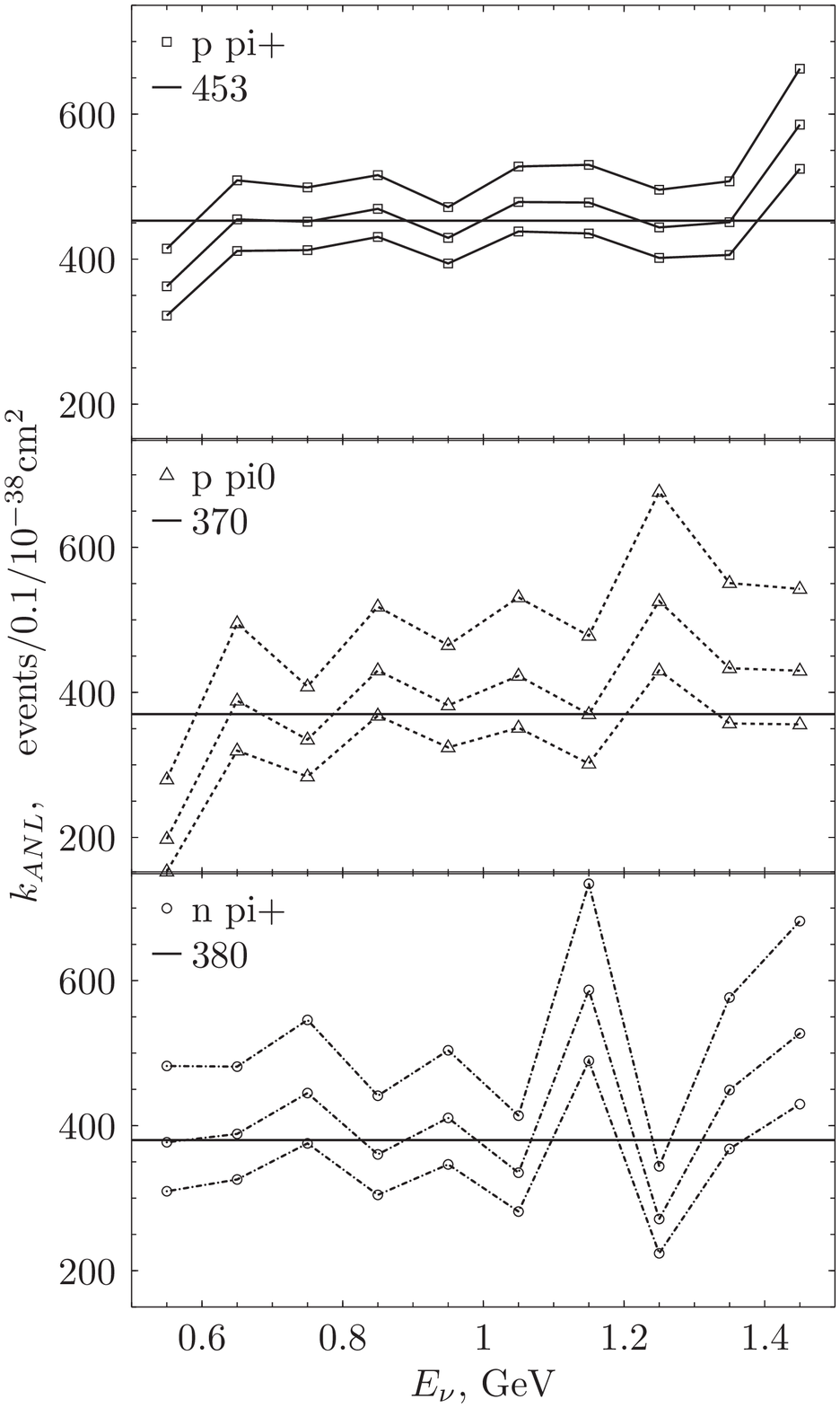}
\caption{The ANL transformation coefficient, determined from the data on neutrino event distributions for various final states.
The three lines reflect an error band obtained from the errors of the published data.}
\label{fig:k-ANL-fit}
\end{figure}

Fig.~\ref{fig:k-ANL-fit} shows the coefficients $k$ determined from Eq.~(\ref{k-fit}) and the ANL data.
The flux is taken from the histogram in \cite{Barish:1977qk}.
The experimental points for the cross sections are interpolated with splines.
The errors for the cross sections are used to estimate the  errorbands of the coefficient.

For each channel the middle curve shows the central value of $k$  which
corresponds to the central value of the cross section. The lower (upper) curves correspond to the
maximal (minimal) values of the cross section  and serve as error bands for $k$.
For each channel $k$ is fitted as a constant value, each point weighted with its
maximal error. The results
\begin{equation}
\begin{array}{l} \di
k_{ANL}^{p\pi^+}=(453\pm 16) \frac{events/0.1}{10^{-38}\cm^2},\
\\[2mm] \di
k_{ANL}^{p\pi^0}=(370\pm 31) \frac{events/0.1}{10^{-38}\cm^2},\
\\[2mm] \di
k_{ANL}^{n\pi^+}=(380\pm 31) \frac{events/0.1}{10^{-38}\cm^2}
\end{array}
\label{k-ANL}
\end{equation}
are shown as straight lines in Fig.~\ref{fig:k-ANL-fit}. The factor $0.1$  comes from the $E_\nu$
binning in Fig.~\ref{fig:k-ANL-fit}.

Comparing $k_{ANL}$ with the corresponding values for each channel separately, we conclude
that an accuracy of around $(453-370)*2/(453+370)\approx 20\%$ should be attributed to it.
This value is consistent with the ANL flux uncertainty estimated in \cite{Radecky:1981fn} as $15\%$.

The procedure considered here is inverse to what
experimentalists do to determine the absolute cross sections. Here its main purpose
is to estimate a reasonable accuracy requirement for fitting
theoretical curves to the data. We conclude, that for the ANL experiment an agreement
within $20\%$ should be considered as perfect.

\subsection{$Q^2$ distribution }

\begin{figure}[!hbt]
\includegraphics[width=\columnwidth]{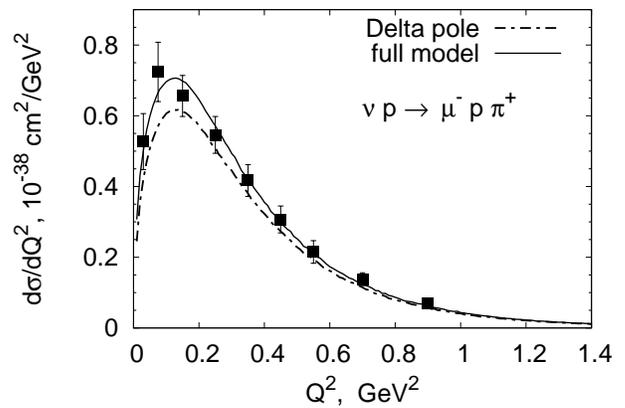}
\caption{Cross section $d\sigma/dQ^2$  averaged over the ANL neutrino energy flux for the final state
$\mu^- p \pi^+$. The integration over $W$ is performed with a
$W<1.4\GeV$ cut, in agreement with the experiment \cite{Radecky:1981fn}. Data are shown as filled squares.}
\label{fig:ANL-distQ2-proton}
\end{figure}

For the $p \pi^+$ channel the $Q^2$-distribution is given by the ANL experiment as an absolute cross section
$d\sigma/dQ^2$ for events with the invariant nucleon--pion mass cut $W<1.4\GeV$ and neutrino energy cut $0.5\GeV<E_\nu<6\GeV$.
Our results, presented in Fig.~\ref{fig:ANL-distQ2-proton}, show a good agreement with the
experimental data.

The same distribution, but without energy cut, as well as those for the $p \pi^0$ and $n \pi^+$ channels,
are presented  as events per $Q^2$ interval. As mentioned before, the axial form factors of the
theoretical model were fitted to the $p\pi^+$ channel of  the ANL data; we thus use
$k_{ANL}\stackrel{!}{=}k_{ANL}^{p\pi^+}$ given in Eq.~(\ref{k-ANL}) as our transformation coefficient for all distributions.
In comparing our theoretical results with the data we therefore normalize them to each other by
multiplying the theory results with the factor $k_{ANL}/{f_{ch}}$ with $f_{ch}$ for the various channels given in Eq.~(\ref{f_ch}).

\begin{figure}[!hbt]
\includegraphics[width=\columnwidth]{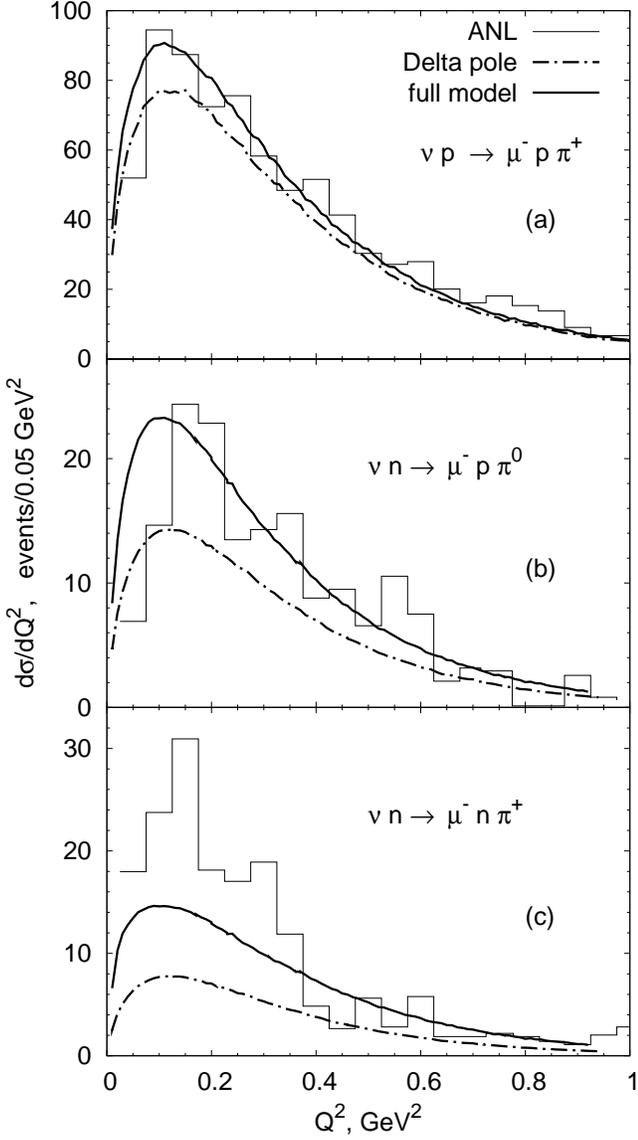}
\caption{Cross section $d\sigma/dQ^2$  averaged over the ANL neutrino energy flux for the final states:
 (a) $\mu^- p \pi^+$,  (b) $\mu^- p \pi^0$, and (c) $\mu^- n \pi^+$ . The integration over $W$ is performed with a
$W<1.4\GeV$ cut, in agreement with the experiment \cite{Radecky:1981fn}. Data are shown as histograms.}
\label{fig:ANL-distQ2}
\end{figure}

Our results are shown in Fig.~\ref{fig:ANL-distQ2} and
compared with the experimental histograms.
The Delta pole contributions to
the $p\pi^0$ and  $n \pi^+$ channels are noticeably below the data. The extra contribution
from the background adds around $50\%$ for the $p\pi^0$ channel, which overshoots the data at
low $Q^2$. The general agreement of our curve with the data is very good.
For the $n\pi^+$ channel, the background contribution adds $100\%$ to the cross section, which, however,
is still not enough to reach the experimentally observed values at low $Q^2$.  This could hint at a contribution
of the higher mass isospin-1/2 resonances, not considered in this work, which may decay into one pion final state.
With the estimated intrinsic uncertainty of $20\%$, the overall agreement should be
considered as good.

\subsection{$W$ distributions}

Next, we present data for invariant mass distributions. In previous theoretical
investigations \cite{Rein:1980wg,Paschos:2003qr,Hernandez:2007qq},
only distributions versus pion-nucleon invariant mass  $W(N \pi)$ were calculated.
The ANL and BNL experimental data are available also for nucleon-muon $W(\mu N)$
and pion-muon $W(\mu \pi)$ combinations. These additional data can be used to constrain the
theory even further.

\begin{figure}[!hbt]
\includegraphics[width=\columnwidth]{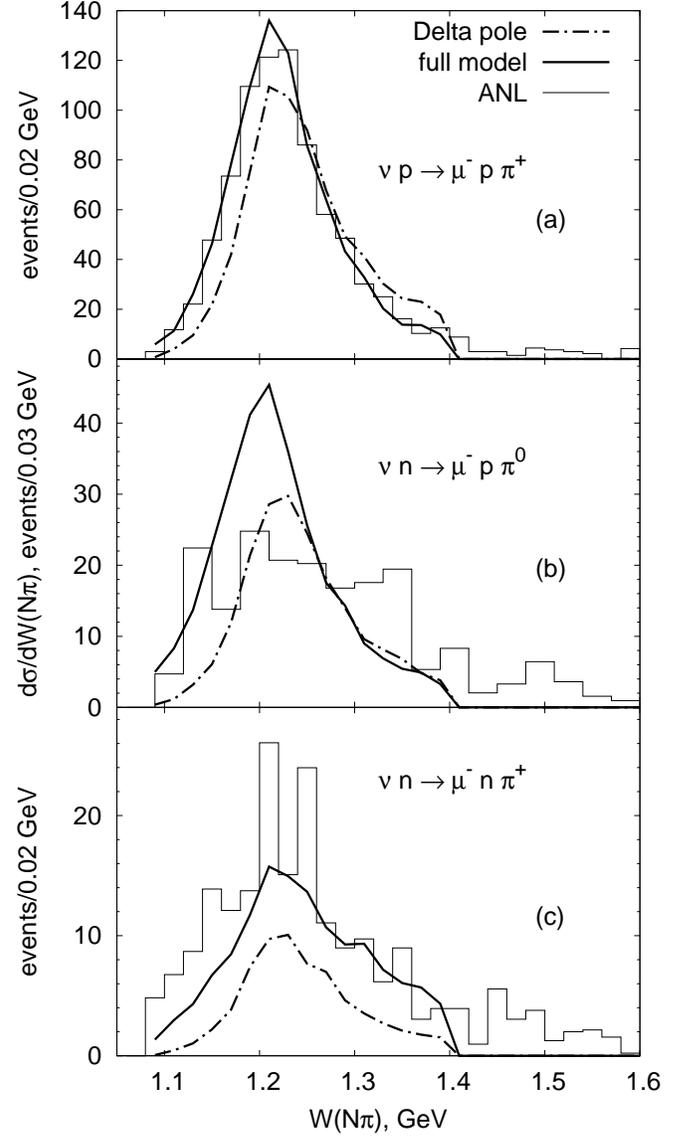}
\caption{The nucleon-pion invariant mass distributions, averaged over the ANL flux.
The full model calculations (solid curve) and Delta pole contribution (dash-dotted curve) are shown.
The experimental data from \cite{Radecky:1981fn}  are shown as histograms.}
\label{fig:ANL-distW-Npi}
\end{figure}

Fig.~\ref{fig:ANL-distW-Npi}a shows the $W(N\pi)$ distribution for the $p\pi^+$ channel. The agreement of our full model calculations
(solid curve) with the histogram is very good.
Of interest is the region of low $W$, near the one pion production threshold.
In this region, the experimental data show a noticeable rise with increasing $W$,
which is in agreement with the full model prediction. The Delta pole contribution (dash-dotted curve), on the other hand,
grows rather slowly.

\begin{figure}[!hbt]
\includegraphics[width=\columnwidth]{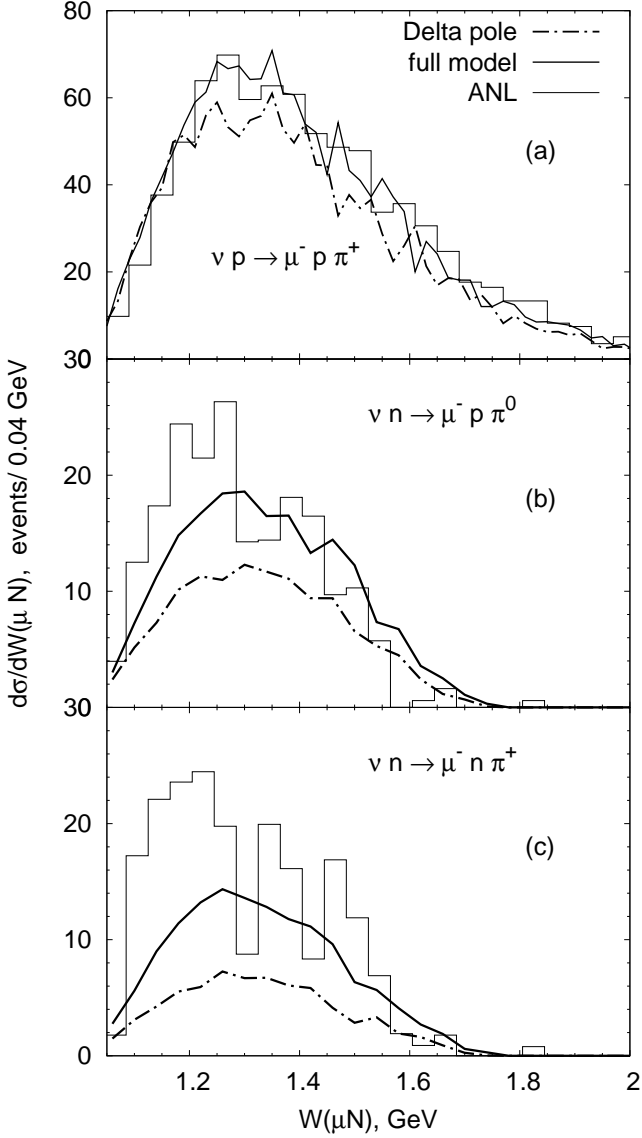}
\caption{The muon-nucleon invariant mass distributions, averaged over the ANL flux.
The full model calculations (solid curve) and Delta pole contribution (dash-dotted curve) are shown.
The experimental data from \cite{Radecky:1981fn}  are shown as histograms.}
\label{fig:ANL-distW-muN}
\end{figure}
For the $p\pi^0$ and $n\pi^+$ final states, as shown in Figs.~\ref{fig:ANL-distW-Npi}b,c,
the agreement of the full model with the histogram is
reasonable. For the $p\pi^0$ channel the full model overestimates events in the Delta peak region
and underestimates them immediately above this peak. For the $n\pi^+$ channel, the data
are underestimated below the Delta peak. 

At low $W$  the background gives a noticeable contribution for both channels,  in line with the data.
For different final states the background contributions above the
Delta peak are very different: small negative for $p\pi^+$, very small for $p\pi^0$ and positive for $n\pi^+$.

\begin{figure}[!hbt]
\includegraphics[width=\columnwidth]{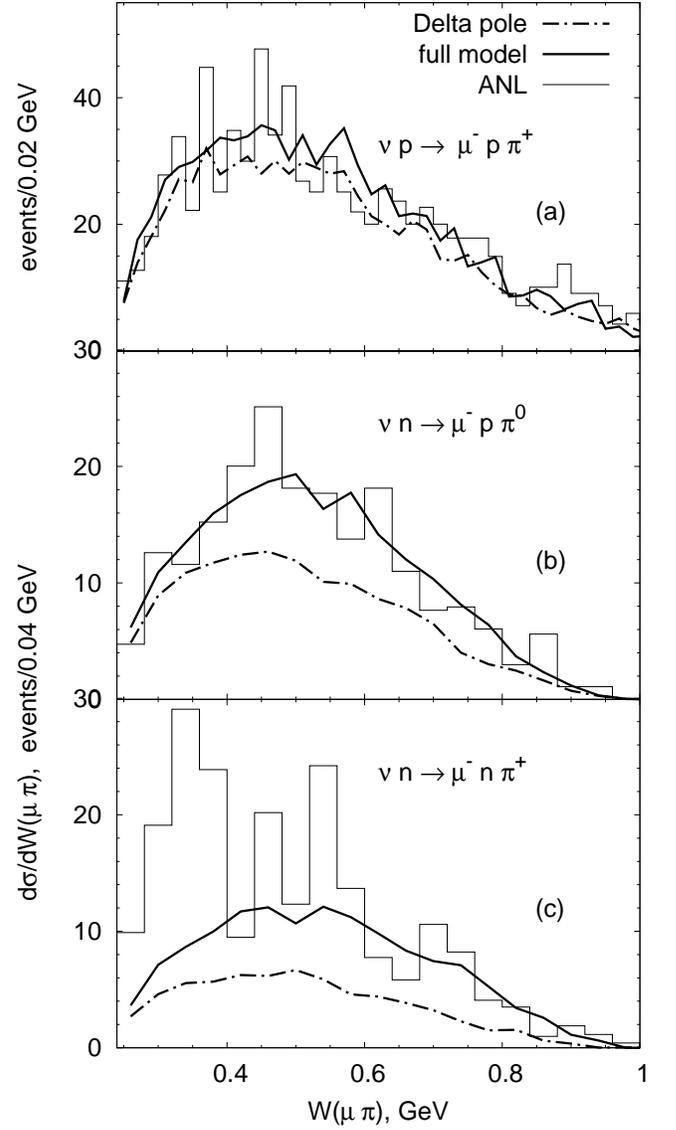}
\caption{The muon-pion invariant mass distributions, averaged over the ANL flux.
The full model calculations (solid curve) and Delta pole contribution (dash-dotted curve) are shown.
The experimental data from \cite{Radecky:1981fn}  are shown as histograms.}
\label{fig:ANL-distW-mupi}
\end{figure}

While the $W(N\pi)$ distributions are mainly sensitive to the $\Delta$ excitation, the distributions
$W(\mu N)$ and $W(\mu\pi)$  test the angular distribution of the $\nu N$ interaction.

The $W(\mu N)$ and $W(\mu \pi)$ distributions shown in Figs.~\ref{fig:ANL-distW-muN}, \ref{fig:ANL-distW-mupi}
also agree reasonably well with our calculations.
Recall, that in the ANL experiment the $p\pi^+$ data are presented for the whole neutrino energy flux,
which only vanishes at $E_\nu=6\GeV$. This can explain the large tail in this distribution.
For the $p\pi^0$ and $n\pi^+$ final states, on the other hand, the experimental data
(as well as our calculations) are limited to $E_\nu<1.5 \GeV$, so that the large $W(\mu N)$
are not kinematically accessible.

The full model and the Delta pole terms give curves of similar form, but different magnitude in the various
isospin channels. For the $p\pi^+$ channel the agreement of our the full model with the data
is very good. In the $p\pi^0$ channel the data are underestimated at low $W(\mu N)$.
For the $n\pi^+$ channel the data are underestimated at both low $W(\mu N)$ and low $W(\mu\pi)$.

The overall results clearly indicate the necessity to include the nonresonant contribution
is addition to that of the Delta pole. 
Even with some underestimation, the full model curves show much better agreement with
the  whole set of data than the Delta pole terms alone.
The background adds around $10\%$ to the  $Dp$ cross section for the $p\pi^+$ channel,
around  $50\%$ for the $p\pi^0$ one and around $100\%$  for the  $n\pi^+$  one.

From  Figs.~\ref{fig:ANL-distW-Npi}--\ref{fig:ANL-distW-mupi}  we conclude,
that within the experimental accuracy available, the data presented can discriminate
between the Delta pole  and the full model curves and are compatible with the HNV background model.

\section{Comparison with the BNL  experiment  \label{BNL}}

\subsection{Transformation from events to absolute cross section}

The results of the BNL experiment on $Q^2$ or $W$ distributions, like those of the ANL experiment,  are
presented  as events.
Data on the distribution of events in neutrino energy are also available \cite{Kitagaki:1986ct} (see Fig. 2 there),
so we use the same procedure as before to determine the transformation coefficient and estimate its accuracy.

The rate correction coefficients $f_{ch}$ for BNL are given in \cite{Kitagaki:1986ct} (see Table II there)
\begin{equation}
 f^{BNL}_{p\pi^+}=1.12, \quad
 f^{BNL}_{p\pi^0}=1.05, \quad  f^{BNL}_{n\pi^+}=0.89 \ .
\label{f_ch_BNL}
\end{equation}

\begin{figure}[!hbt]
\includegraphics[width=\columnwidth]{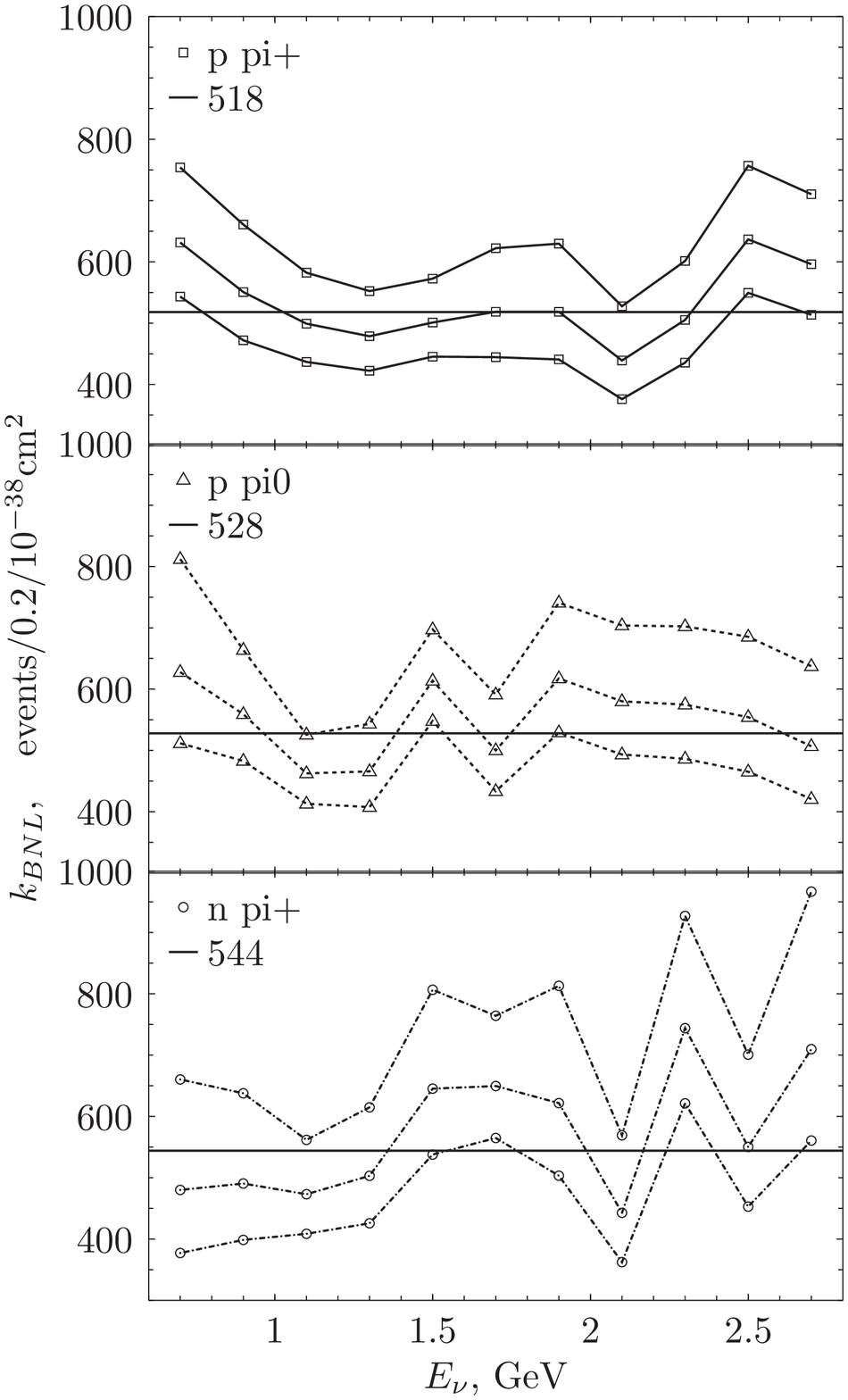}
\caption{The BNL transformation coefficient, determined from the data on neutrino event distributions for various final states.
The three lines reflect an error band obtained from the errors of the published data.}
\label{fig:k-BNL-fit}
\end{figure}

Fig.~\ref{fig:k-BNL-fit} shows the coefficient $k$ determined from Eq.~(\ref{k-fit}) and the BNL data.
The flux is calculated from the observed quasielastic events and given in \cite{Baker:1981su}.
The experimental points for the cross sections are interpolated with splines.
The error bars of the cross sections are used to estimate the  error bands of the coefficient.

For each channel, the middle curve with data points on it shows the central value of $k$, while
the lower (upper) curves serve as error bands for $k$.
For each channel,  $k$ is fitted as a constant value, each point weighted with its
maximal error. The results
\begin{equation}
\begin{array}{l} \di
k_{BNL}^{p\pi^+}=(518\pm 29) \frac{events/0.2}{10^{-38}\cm^2},\
\\[2mm] \di
k_{BNL}^{p\pi^0}=(528\pm 30) \frac{events/0.2}{10^{-38}\cm^2},\
\\[2mm] \di
k_{BNL}^{n\pi^+}=(544\pm 42) \frac{events/0.2}{10^{-38}\cm^2}
\end{array}
\label{k-BNL}
\end{equation}
are shown as straight lines in Fig.~\ref{fig:k-BNL-fit}.

Comparing $k_{BNL}$ for the three channels, we attribute an
accuracy of around $(544-518)*2/(544+518)=5\%$ to $k_{BNL}$.
We conclude, that when comparing with the BNL experiment,
an agreement within $5\%$ should be considered as perfect.

\subsection{$Q^2$ distribution}

The data on the $Q^2$ distribution for the $p \pi^+$ channel are presented in \cite{Kitagaki:1986ct} for
the neutrino energy cut $0.5 \GeV<E_\nu<6\GeV$ and the invariant mass cut $W<1.4 \GeV$.
Since the latter corresponds to the range of applicability of the HNV model, we can
normalize the area under the full model theoretical curve to that under the experimental data.
This is another way to estimate the transformation coefficient  $k^{Q2}_{BNL}$ [which for a perfect experiment
must be equal to those in Eq.~(\ref{k-BNL})]:
\begin{equation}
\begin{array}{c}
\left( \frac{d \sigma}{d Q^2}\right)_{exper} = k_{BNL} \cdot  \left(\frac{d \sigma}{d Q^2} \right)_{theor},
\\[2mm]
 k^{Q2}_{BNL} = 182.5\frac{events/ 0.05 \GeV^2}{10^{-38} cm^2 / \GeV^2 } = 730\frac{events/ 0.2}{10^{-38} cm^2}   \ .
\end{array}
\label{k-Q2-BNL}
\end{equation}
This value exceeds the typical value of $k_{BNL}$ determined in Eq.~(\ref{k-BNL}) by more than $30\%$
(even more for the $p\pi^+$ channel), and thus cannot be considered as consistent.

Speculating about the possible origin of this inconsistency, we note that
the values (\ref{k-BNL}) for different channels agree among themselves quite well, which may hint at the
consistent treatment of the rate correction coefficients $f_{ch}$. Keeping also in mind, that the flux is
the same for various channels, a possible  way to explain the above inconsistency would be to suppose
that the cross sections are overestimated by $30\%$. This is exactly the difference between ANL and BNL integrated
cross sections. Thus, by reducing $\sigma_{tot(ch)}^{BNL}$ by $30\%$ one would simultaneously reach agreement with ANL
and obtain $k^{ch}_{BNL}$ consistent with $k^{Q2}_{BNL}$.

With the data as they are, we reestimate a realistic uncertainty as $30\%$.

Thus, for the BNL experiment we aim at shape-only comparison and
hereinafter use $k^{Q2}_{BNL}$ (\ref{k-Q2-BNL}) as our transformation coefficient.

\begin{figure}[!hbt]
\includegraphics[angle=-90,width=\columnwidth]{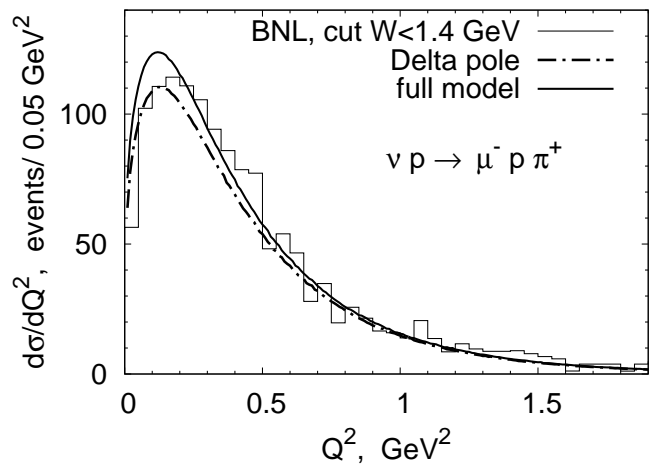}
\caption{Cross section $d\sigma/dQ^2$  averaged over the BNL neutrino energy flux for the final state
$\mu^- p \pi^+$. The integration  is performed with the
$W(N\pi)<1.4\GeV$ cut, in agreement with the experimental data \cite{Kitagaki:1986ct}, which are shown as histogram.}
\label{fig:BNL-distQ2-proton}
\end{figure}

The comparison of the $Q^2$ distribution with our calculations is shown in Fig.~\ref{fig:BNL-distQ2-proton}.
For the $p \pi^+$ channel, as expected, the background gives only a small contribution in addition to the leading Delta pole term.
In agreement with earlier calculations \cite{Lalakulich:2005cs},
the peak of the curve is located in the region $Q^2=0.1-0.12 \GeV^2$ and is shifted with respect to the data peak
at $Q^2=0.18-0.2 \GeV^2$. This disagreement has been known for a long time with similar results obtained within
various models \cite{Kitagaki:1986ct,Lalakulich:2005cs,Graczyk:2007xk,Leitner:2009zz} and is not resolved.
Inclusion of the background does not change the peak position.

\begin{figure}[!hbt]
\includegraphics[width=\columnwidth]{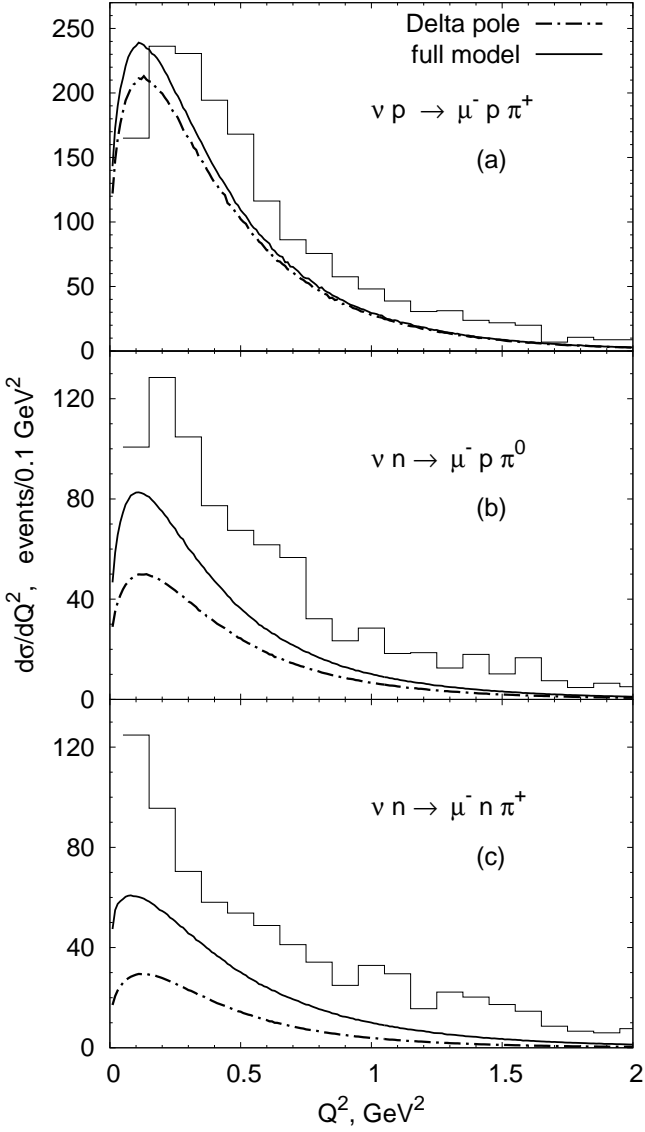}
\caption{The $d\sigma/dQ^2$ cross section averaged over the BNL neutrino energy flux for the final states:
 (a) $\mu^- p \pi^+$,  (b) $\mu^- p \pi^0$, and (c) $\mu^- n \pi^+$. The integration is performed with the
$W(N\pi)<1.4\GeV$ cut, corresponding to the range of applicability of the HNV model .
The experimental data \cite{Kitagaki:1986ct} shown as histograms are without $W$ cut.}
\label{fig:BNL-distQ2}
\end{figure}

The data for the other channels are given in \cite{Kitagaki:1986ct} without
any cut on $W$, and for the whole neutrino flux (that is $0.34\GeV < E_\nu < 6\GeV$)
as events per $0.1 \GeV^2$ interval. They are compared with our calculations
in Fig.~\ref{fig:BNL-distQ2}, where the same transformation coefficient
$k_{BNL}^{Q^2}/f^{BNL}_{ch}$ [Eq.~(\ref{k-Q2-BNL})] is used.
The same coefficient is also used further for various $W$ distributions.

For the $p \pi^+$ channel, as expected, our full model calculations (solid line)
is below the experimental histogram.
The area under our curve is $15\%$ below the area under the histogram.
Since the data include higher invariant masses whereas our calculations contain a cutoff of $1.4\GeV$,
corresponding to the range of validity of the HNV model, this implies that
$15\%$ of all events in this channel should be attributed to the
higher mass isospin-3/2 resonances, such as $P_{33}(1600)$,
$S_{31}(1620)$, $D_{33}(1700)$, and their interferences with the background,
which are not considered here.

For the $p\pi^0$ and $n\pi^+$ channel the background significantly increases the cross section in comparison
with the Delta pole contribution. However, our full model curves  are still much lower than the histograms, which indicates
a large contribution of higher mass isospin-1/2 and -3/2 resonances and their interferences. The relative importance of these
events is estimated by comparing the areas under the theoretical curve and experimental histogram, as it was
described for the $p\pi^+$ channel, and it appears to be $43\%$ for $p\pi^0$ and $46\%$ for $n\pi^+$.
This will also be demonstrated further in the $W(N\pi)$ invariant mass distribution.

Notice also, that in the $p\pi^0$ channel [Fig.~\ref{fig:BNL-distQ2}(b)] the peak of our curve is shifted to the left 
with respect to the histogram. This effect is the same as in $p\pi^+$ channel in Fig.~\ref{fig:BNL-distQ2-proton}, 
but it is revealed here  with less significance because of the larger $Q^2$  binning.

\subsection{$W$ distribution}

\begin{figure}[hbt]
\includegraphics[width=\columnwidth]{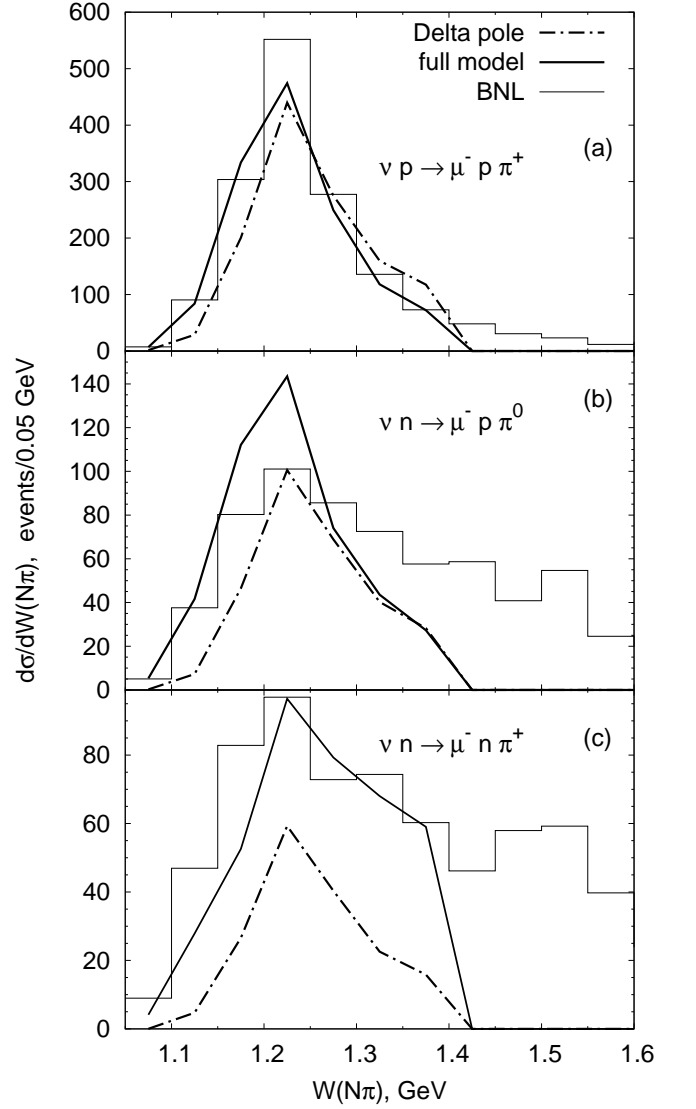}
\caption{The nucleon-pion invariant mass distributions, averaged over the BNL flux.
The full model calculations (solid curve) and Delta pole contribution (dash-dotted curve) are shown with the
cut $W(N\pi)<1.4 \GeV$.
The experimental BNL data \cite{Kitagaki:1986ct} which do not contain this cut  are shown as histograms.}
\label{fig:BNL-distW-Npi}
\end{figure}

Now we proceed with calculating the invariant mass distributions. Fig.~\ref{fig:BNL-distW-Npi} shows the
$W(N\pi)$ distribution for the three final states. Our calculations are done only up to $W<1.4\GeV$,
which is the range of applicability of the HNV model, while the experimental data are available also for
higher $W$.

For the $p\pi^+$ channel, our calculations show a very good agreement with the data. In the region $1.05 \GeV< W< 1.4 \GeV$
the area under our curve coincides with the area under the histogram with an accuracy better than $1\%$.

To estimate how many events belong to the high $W(N\pi)$ region, we calculate the area under the histogram with the
cut $W(N\pi)<2.0\GeV$ and compare it to that with the cut $W(N\pi)<1.4\GeV$. We find that $\gtrsim 10\%$
of all events belong to the $W(N\pi)>1.4 \GeV$ region. This is in agreement with the conclusion previously derived from the $Q^2$ distributions.

For both the $p\pi^0$ and $n\pi^+$ channels, the background is essential, especially at very low $W(N\pi)$,
where it significantly increases the cross sections.
The overall agreement of our calculations with the data is reasonable.
Similar to our comparison with the ANL experiment, for the $p\pi^0$ channel the full model overestimates the data
in the region of Delta peak and underestimates them immediately above this peak. For the
$n\pi^+$ channel, the data are underestimated below the Delta peak.
The percentage of events with $W(N\pi)>1.4 \GeV$ is estimated to be
$45\%$ for the $p\pi^0$ and  $44\%$ for the $n\pi^+$,
which is in good agreement with the previous results obtained when discussing the $Q^2$ distribution.

As in our calculations for the ANL experiment, in the $W$ region above the Delta peak the background
contribution is different for  different channels: negative for $p\pi^+$, small for $p\pi^0$ and positive for $n\pi^+$.
The comparison of the two latter channels shows that the effective background $d\sigma^{bgr}/dW(N\pi)$ as it is described within the HNV model
does not support the isospin-1/2 hypothesis.

\begin{figure}[hbt]
\includegraphics[width=\columnwidth]{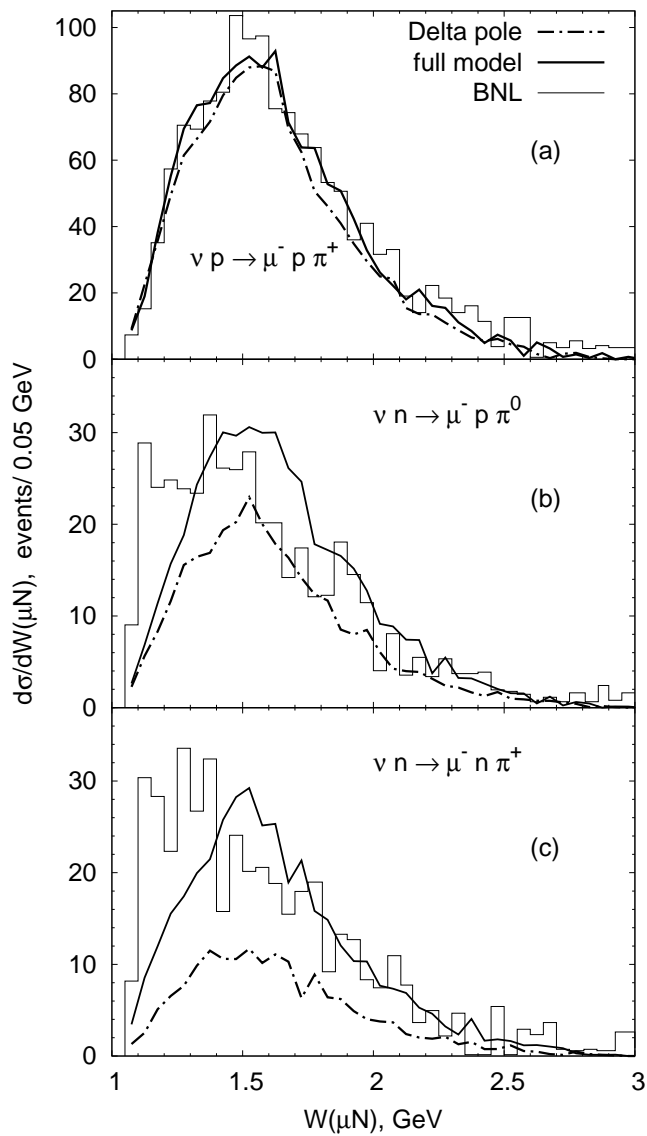}
\caption{The muon-nucleon invariant mass distributions, averaged over the BNL flux, for the events with $W(N\pi)<1.4\GeV$
compared to the BNL data \cite{Kitagaki:1986ct} shown as histograms.}
\label{fig:BNL-distW-muN}
\end{figure}

Fig.~\ref{fig:BNL-distW-muN} shows the muon--nucleon invariant mass distribution compared with the BNL data subject to
selection $W(N\pi)<1.4\GeV$.

For the $p\pi^+$ channel, the background  slightly increases the cross section.
For the $p\pi^0$ and $n\pi^+$ channels it gives a significant contribution,
which noticeably improves agreement with the histograms.
For the $p\pi^0$ channel, the data show an excess of events over our curve at low $W(\mu N)$.
As explained in \cite{Kitagaki:1986ct},
''this excess comes in part from the misidentified $\nu n\to \nu \pi^- p$ and $nn\to np\pi^-$ events which belong to the experimental
background`` not subtracted from the data.

This generally means that the experimental background is nonuniform and requires a more detailed treatment.
Indeed, all experimental histograms show the distributions of observed (raw) events. The
rate correction factors $f_{ch}^{BNL}$ are provided experimentally as constant factors that do not
depend on kinematics. In this way, we correct the total number of events, but
the nonuniformity of the experimental corrections remains unaccounted for and can reveal itself in
all distributions.

For the $n\pi^+$ channel, the agreement is good in the region of high $W(\mu N)$,
but again we underestimate the data for low $W(\mu N)$.

\begin{figure}[htb]
\includegraphics[width=\columnwidth]{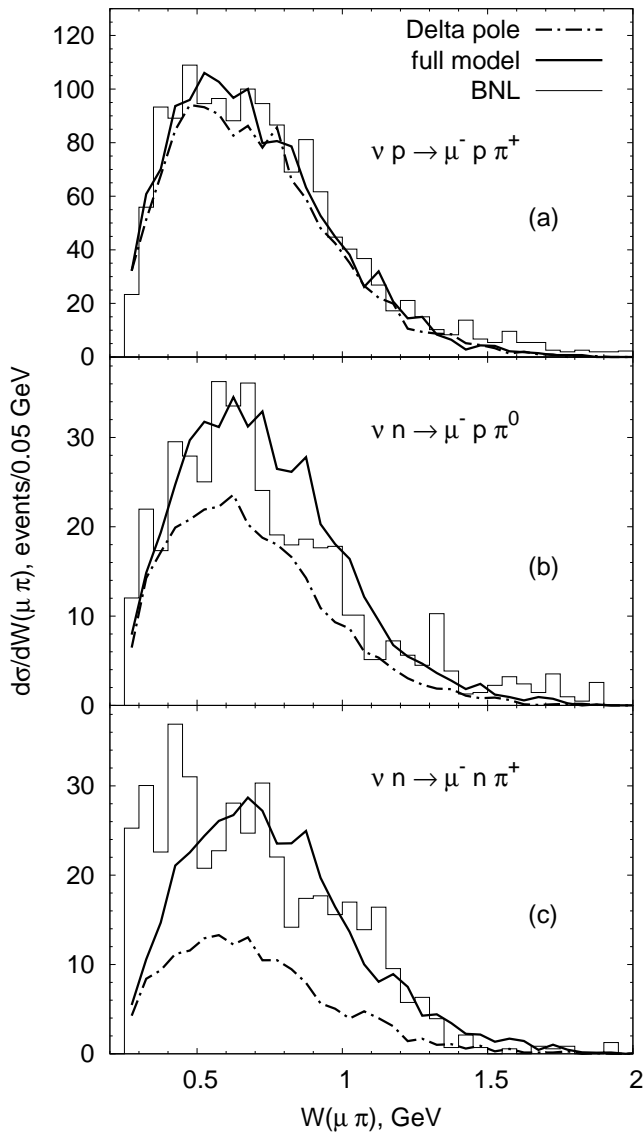}
\caption{The muon-pion invariant mass distributions, averaged over the BNL flux, for the events with $W(N\pi)<1.4\GeV$
compared to the BNL data \cite{Kitagaki:1986ct}  shown as histograms.}
\label{fig:BNL-distW-mupi}
\end{figure}

Fig.~\ref{fig:BNL-distW-mupi} shows the muon-pion invariant mass distribution compared with the BNL data subject to
selection $W(N\pi)<1.4\GeV$. The agreement of our calculations with the data is good for the $p\pi^+$ and $p\pi^0$ channels,
while for the $n\pi^+$ one we underestimate the data at low $W(\mu\pi)$.

All the three $W$ distributions taken together clearly show the importance of background terms in reaching a
reasonable description of  the data for all the three  invariant mass  distributions considered.
Delta pole term alone, showing a good agreement with the data for the $p\pi^+$
channel, significantly underestimates all the data for the $p\pi^0$ and $n\pi^+$ channels.

Some discrepancies between the full model calculations and data at low $W$s [for all three $W(N\pi)$, $W(\mu N)$ and $W(\mu\pi)$]
for both ANL and BNL may point to the necessity to improve the model.
As discussed earlier in this section, the BNL experimentalists themselves attributed the excess of
events in $W(\mu N)$ distribution of the $p\pi^0$ channel to some misidentified events. Similar
excess in other channels for low $W$ may have a similar origin.

\section{Discussion and conclusion}

In Sec.~\ref{model}  we have outlined the HNV model and the phenomenological form factors for various diagrams.
As ensured by comparison with electroproduction in Sec.~\ref{electron} the model is applicable up to
nucleon-pion invariant mass of $W(N\pi)<1.4\GeV$ and provides about the same level of accuracy as the MAID model.

For neutrino reactions, as discussed in Sec.~\ref{neutrino}, the model predicts a background contribution,
which is small for the $p\pi^+$ channel (at the level below $10\%$
for neutrino energies above $1 \GeV$), at the level of $30\%$ (with respect to the full model cross section, that is $50\%$
with respect to the Delta pole contribution) for the  $p\pi^0$ channel and at the level of $50\%$ (with respect to the full
model cross section, that is $100\%$ with respect to the Delta pole contribution)  for the  $n\pi^+$ channel.
The effective background does not satisfy the isospin-1/2 hypothesis.

The HNV model describes the available data set on neutrino and antineutrino reactions on nucleons
reasonably well, with an accuracy that approximately corresponds to the accuracy with which different data agree among themselves.

For neutrinos the absolute values of the integrated cross section are available from the ANL and BNL experiment,
with the BNL data being systematically higher.  The Delta axial form factors were fitted in \cite{Hernandez:2007qq}
to the ANL integrated cross section and the $Q^2$ distribution for the $p\pi^+$ channel so that the full model agrees with them by definition.
Agreement with the ANL data for the $p\pi^0$ and $n\pi^+$ channels shows that the model gives a good description of
the background. The BNL data lie around $30\%$ higher for all channels.

For antineutrinos only very few data are available from the Gargamelle experiments. The agreement of the full model calculations
with the data is good for the $p\pi^-$ channel and overestimates the data for the $n\pi^-$ one.

For most of the differential cross sections available experimentally no information about the absolute
value of the cross section is available; the data are presented as raw events per $Q^2$ or $W$ interval.
By estimating the transformation coefficients for the ANL and BNL experiments in Secs.~\ref{ANL}a,\ref{BNL}a,
we have shown that the realistic accuracy for the ANL experiment is around $20\%$,  with the uncertainties
coming from the inconsistencies in the various channels.
For the BNL experiment the various channels agree within $5\%$; here the overall accuracy about $30\%$
comes from the disagreement of the $Q^2$ distribution, discussed in Sec.~\ref{BNL}b,
with the integrated event distribution.

As discussed in Secs.~\ref{ANL},\ref{BNL}, the overall agreement with the data
is perfect for the $p\pi^+$ channel and reasonable for the $p\pi^0$ and $n\pi^+$ ones.
For the latter two channels the full model calculations systematically underestimate the
experimental histograms at low $W$s. These discrepancies 
may hint at the necessity to improve the model, but they may as well come from the recognized
nonuniformity of the experimental background.

Even without absolute normalization, and even taking into account the discrepancies observed,
all the ANL and BNL data taken together are able to discriminate between the full model and
the leading Delta pole contribution. When all three final states accessible for neutrino reactions
($p\pi^+$, $p\pi^0$, $n\pi^+$) are considered, the data definitely demand the nonresonant background and
favor the full model calculations.

\acknowledgements
This work is supported by DFG. The authors are grateful to M. Valverde for providing us with
his unpublished results for comparison.

\bibliographystyle{apsrev}
\bibliography{nuclear.bib}

\end{document}